\documentclass[12pt,preprint]{aastex}

\newcommand{\degree}{\ensuremath{^\circ}}
\providecommand{\e}[1]{\ensuremath{\times 10^{#1}}}

\slugcomment{To be submitted to the Astronomical Journal}

\shorttitle{Asymmetric Thick Disk II}
\shortauthors{Larsen et al.}

\begin{document}


\title{Mapping the Asymmetric Thick Disk: II \\ Distance, Size and Mass of the
Hercules Thick Disk Cloud}


\author{Jeffrey A. Larsen}
\affil{Physics Department, United States Naval Academy,
    Annapolis, MD 21402}
\email{larsen@usna.edu}

\author{Juan E. Cabanela}
\affil{Department of Physics and Astronomy, Minnesota State University Moorhead, Moorhead MN, 56563}
\email{cabanela@mnstate.edu}

\and

\author{Roberta M. Humphreys}
\affil{Astronomy Department, University of Minnesota, Minneapolis MN, 55455}
\email{roberta@umn.edu}




\begin{abstract}
The Hercules Thick Disk Cloud \citep{lar08} was initially discovered as an excess in the number of faint
blue stars between quadrants 1 and 4 of the Galaxy.
 The origin of the Cloud could be an interaction
with the disk bar, a triaxial thick disk or a merger remnant or stream. To better map the spatial extent of the
Cloud along the line of sight, we have obtained multi-color UBVR photometry for 1.2 million stars in 63 fields 
approximately 1 square degree each. Our analysis of the fields beyond the apparent
boundaries of the excess have already ruled out a triaxial thick disk as a likely explanation
\citep{lar10}. In this paper
we present our results for the star counts over all of our fields, determine the spatial extent of the over density across and  along the line of sight,  and
estimate the size and mass of the Cloud.  Using photometric parallaxes, the stars responsible for
the excess are between 1 and 6 kiloparsecs from the Sun, 0.5 -- 4 kpc above the
Galactic plane, and extends approximately 3-4
kiloparsecs across our line of sight. It is thus a major substructure in the
Galaxy. The distribution of the excess along our sight lines corresponds with the density contours of the bar in the Disk, and its most distant stars are  directly over the bar.
We also see through the Cloud to its far side. Over the entire 500 square degrees of sky 
containing the Cloud, we estimate more than 5.6 million stars 
and 1.9 million solar masses of material.  If the over density is associated with the bar, 
it would exceed 1.4 billion stars and  more than than 50 million solar masses.  
Finally, we argue that the Hercules-Aquila Cloud \citep{bel07} is actually the Hercules Thick Disk Cloud.

\end{abstract}



\section{Introduction}

Studies of both stars and gas in the Galaxy are revealing significant 
structure and asymmetries in its motions and spatial distributions.  Some
examples of recent structure include the bar of
stars and gas in the Galactic bulge (\citealt{1991ApJ...379..631B},
\citealt{1994ApJ...429L..73S}), the evidence from infrared surveys for a larger stellar bar in the inner
disk (\citealt{1992ApJ...384...81W}, \citealt{1997MNRAS.292L..15L},                \citealt{2005ApJ...630L.149B}),
the outer ring \citep{2003ApJ...588..824Y},  the discovery of the Sagittarius
dwarf
\citep{1994Natur.370..194I,1995MNRAS.275..591I} and a significant asymmetry of unknown 
origin in the distribution of faint blue stars in Quadrant 1 (Q1) of the inner Galaxy 
\citep{lar96}. Each of these observations provides a significant clue
to the history of the Milky Way. When combined with the growing evidence 
for Galactic mergers in addition to the Sagittarius dwarf, i.e.
 the Monoceros stream (\citealt{2002ApJ...569..245N}, 
\citealt{2003MNRAS.340L..21I}), the Canis Major merger remnant 
(\citealt{2004MNRAS.355L..33M}),
the Virgo  stream (\citealt{2001ApJ...554L..33V}, 
\citealt{2005ApJ...633..205M})
and the recent Hercules--Aquila cloud \citep{bel07}, we now realize that 
the
structure and evolution of our Galaxy have been significantly altered by 
mergers
with other systems.  Indeed the population of the Galactic Halo
and possibly the Thick Disk as well, may be dominated by mergers with 
smaller systems.

Larsen and Humphrey's asymmetry involves faint bluer stars in Quadrant 1 
(Q1) of the inner Galaxy ($l = 20\degree - 45\degree$ at intermediate latitudes)
characterized by an overdensity of $\approx$ 30\% when
compared with complementary longitudes in the Quadrant 4 (Q4 ,
$l = 315\degree - 340\degree$).
The initial discovery was made 
using star counts from the Minnesota Automated Plate Scanner Catalog of the
POSS I (MAPS, \cite{cab03} \footnote{http://aps.umn.edu}).
A more spatially complete survey \citep{par03} with 40
contiguous fields above and below the plane in Q1 and in Q4 above the plane
confirmed the star count excess and found that the asymmetry in Q1, 
while somewhat irregular in shape, was also fairly uniform and covered 
several hundred square degrees in
Q1. It is therefore a major substructure in the Galaxy due to more than small 
scale clumpiness. 
The stars responsible for
the excess were probable Thick Disk stars typically 1 -- 2 kpc from the Sun.
\cite{par04} also found an associated kinematic signature.  The Thick Disk stars in
Q1 have a much slower effective rotation rate $\omega$, compared to the
corresponding Q4 stars, with  a significant lag of 80 to 90 km s$^{-1}$ in
the direction of Galactic rotation, greater than the expected lag of 30 -- 50 km s$^{-1}$ for the Thick Disk population. The asymmetry is now designated the Hercules Thick Disk Cloud \citep{lar08} (hereafter the Hercules Cloud).

The release of the SDSS Data Release 5 (DR5) photometry in the same 
direction as the observed asymmetry in Q1 led to the discovery of 
another feature at much fainter magnitudes, the so-called Hercules-Aquila
cloud \citep{bel07}, however we suggest (\S {4.3} that the over density is actually closer and at the same distances as the nearer Hercules Cloud. 
A second analysis \citep{jur08} of the same dataset and 
led to the confirmation of our nearer Hercules Cloud at its approximate 
distances, though it was initially attributed to a possible stellar ring 
above the plane.  Our 
comparison of the
stellar density distributions in 
Q1 and Q4 above the plane  \citep{lar08} demonstrated
that the excess is in Q1 only and is therefore not consistent with a ring.

With the increasing evidence
for Galactic mergers \citep{1994Natur.370..194I,
1995MNRAS.275..591I, 2002ApJ...569..245N, 2003ApJ...596L.191N,
2004MNRAS.355L..33M, 2005ApJ...633..205M, 2003ApJ...588..824Y, 2006ApJ...639L..13W},
we now realize that the
population of the Galactic Halo, and possibly the Thick Disk as well, may be dominated by mergers with smaller systems.
The Hercules Cloud has no spatial overlap with the path of the
Sagittarius dwarf through the Halo \citep{2001ApJ...547L.133I}, and the predicted
path of the Canis Major dwarf \citep{2004MNRAS.355L..33M}, so its  association  with
either of these well-studied features is unlikely.  Our line of sight
to the asymmetry is also interestingly in the same general direction as the stellar
bar in the Disk
\citep{1992ApJ...384...81W, 1997MNRAS.292L..15L, 2000MNRAS.317L..45H, 2005ApJ...630L.149B}, but the  bar is approximately 5 kpc from the Sun  in this direction. Thus the 
stars showing the excess were mostly between the Sun and the bar, not directly above it. However the maximum extent of the  star count excess
along our line of sight was  not known.

Interpretation of the Hercules Cloud is not clear-cut.
While it might
well be the fossil remnant of a merger,
the star count excess is  also consistent with 
a triaxial Thick Disk or inner Halo as well as a dynamical 
interaction with the stellar bar especially given the corresponding 
asymmetry in the kinematics \citep{par04}.  A rotating bar in the Disk could induce a 
gravitational ``wake'' that would trap and pile up stars behind it 
\citep{1992ApJ...400...80H, 1998ApJ...493L...5D}.  Thus in response to the 
bar, there would not only be an  excess of stars in Q1 over Q4, but 
those stars may show a measurable lag in their rotational velocities as 
observed in Q1.  While similar, a triaxial Thick Disk could also yield different 
effective rotation rates because of  noncircular streaming motions along 
its major axis.

If the Thick Disk is triaxial, we would expect to observe
the star count excess out to greater longitudes, but
it appears to terminate near {\it l} $\sim$ 55$\degree$ \citep{par03}. 
To search for the asymmetry at greater longitudes from the Galactic center, our 
Paper I \citep{lar10} extended the star counts to fainter magnitudes, 
corresponding to greater distances.  Our results 
do not support the triaxial interpretation of the asymmetry. 
We find a statistically
significant excess of faint blue stars for the two innermost Q1 fields at
 {\it l} of 45$\degree$ and 50$\degree$, but the fields at the greater longitudes
 (55$\degree$, 60$\degree$, 65$\degree$ and 75$\degree$) show no significant
 excess including the faintest magnitude intervals.

One of the greatest uncertainties concerning the nature of the Hercules Cloud is its spatial 
extent along the line of sight.
Our earlier work \citep{lar96,par03,par04} used photographic data having completeness
limits of 18 -- 18.5 mag.  To further explore its  possible origins, we
have  mapped the  extent of the spatial asymmetry
to greater distances as a function of Galactic longitude and latitude.
In Paper I we described our CCD observing
program to much fainter limiting magnitudes. In the next section we present a
brief summary of the observations and the data. In \S {3} we describe our analysis of
the star counts and population separation.
The resulting map of the star count excess from photometric parallaxes and the 
size and mass of the
Hercules Cloud are presented in \S {4} and in the last section the
implications for the origin of the asymmetry and the Hercules Cloud
are discussed.

\section{The Observations}

\begin{figure}
\epsscale{0.8}
\plotone{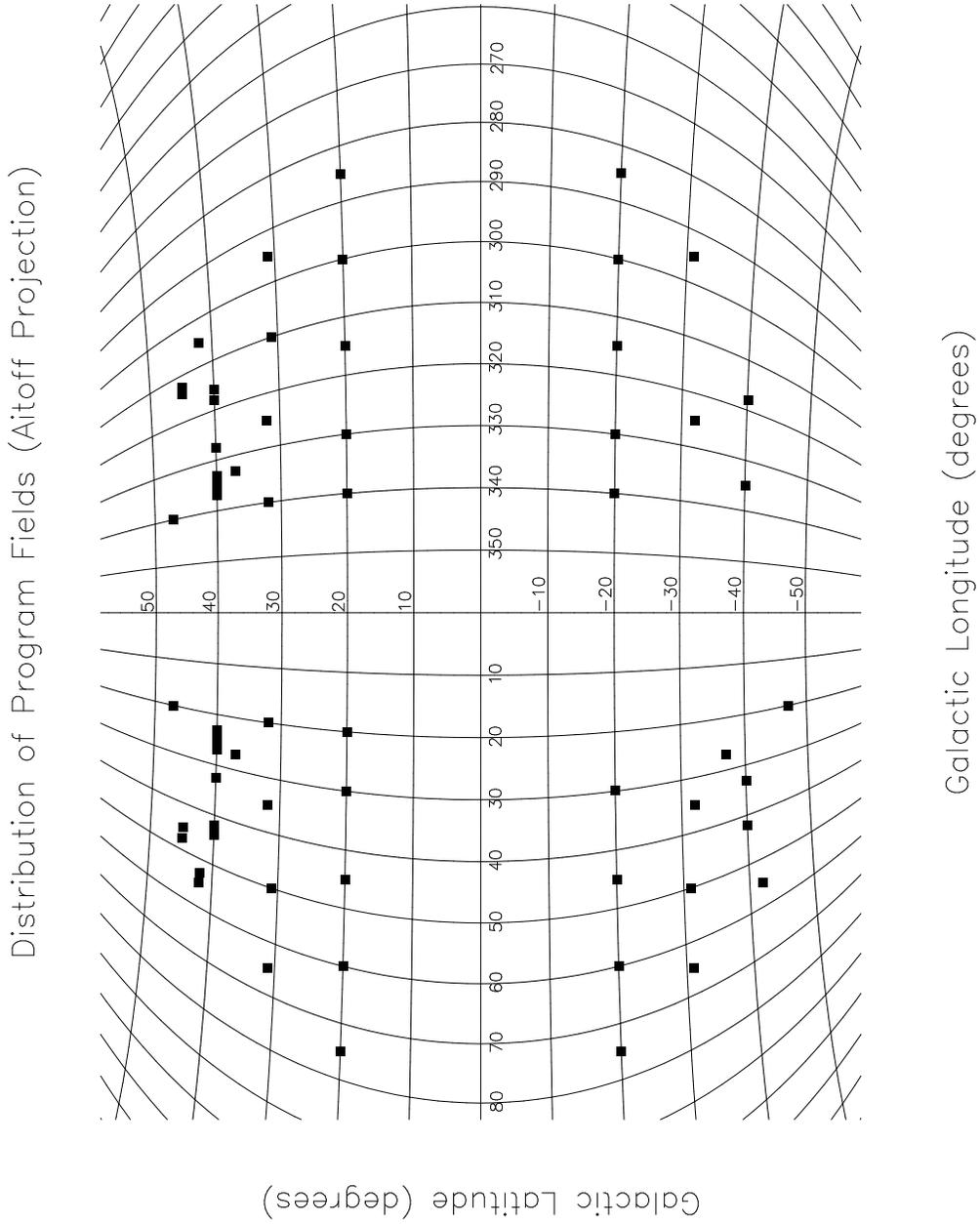}
\vspace{0.5in}
\caption{Distribution of program CCD fields across the Galactic sky. Each dot is roughly the same are of one of our program fields, 1 square degree.\label{fig1}}
\end{figure}

Between 2006 and 2008, we obtained multicolor UBVR CCD images for 67 fields
ranging in longitude from {\it l} = 20$\arcdeg$ to
75$\arcdeg$, and {\it l} = 340$\arcdeg$ to 285$\arcdeg$ and in latitude
from {\it b} = $\pm$20$\arcdeg$ to $\pm$45$\arcdeg$. The total survey covers
47.5 square degrees and includes 1.2 million stars. The distribution of the
program fields on the sky is presented in Figure~\ref{fig1}. These observations were
obtained with the 90Prime camera \citep{2004SPIE.5492..787W} on the
Steward Observatory Bok 90-inch and the
Y4KCam at the 1.0 meter SMARTS telescope at CTIO.
We used a Johnson U,B,V + Cousins-Kron R filter set on both instruments with integration times
selected to reach limiting magnitudes of fainter than 22nd magnitude in V for the 90Prime 
fields and 20th magnitude
for the Y4KCam fields. Field sizes were to be 1 square degree.  In practice, our
results varied.  For the Y4KCam, 
the requirement of 9 sets of images per square degree were not always met. For 90Prime,
the later observations were plagued by electronic problems which 
caused a gradual reduction in the usable imager surface and electronic noise which greatly decreased
the sensitivity to faint objects.
For 90Prime images, saturation of the image in V from long exposures also caused many stars 
brighter than 16 to have incorrect colors (trending towards bluer B-V colors) in many fields, and
were therefore removed from the catalogs. 

A re-reduction of the data for this analysis found that three 90Prime fields were not usable.  A fourth field, while several degrees away from the SMC, 
included many stars from the SMC Extension and  was not used. This leaves 
the 63 fields in Table~\ref{tbl1}.
Complete information on all of the program fields, the CCD reduction techniques, star-galaxy
discrimination, completeness limits and astrometry can be found in Paper I.

Seventeen of our fields overlap some portion of SDSS DR7 \citep{2009ApJS..182..543A}.  
We transformed the SDSS $g'$, $r'$ and $i'$ magnitudes to the Johnson-Cousins Kron system
using the equations of \citet{2006AJ....132..989R} and \citet{2005AN....326..321B} for the
samples of stars found in the overlaps between the two catalogs.   The zero point results of the comparison are summarized in Table~\ref{tbl2}. No significant scale errors were determined and our 
photometric zero points agree to within 0.04 dex.  Individual fields have a scatter in zero points of 0.05 dex. 

The electronic catalogs of all 63 fields listed in Table~\ref{tbl1} are 
available at two websites, \url{http://aps.umn.edu} and \url{http://iparrizar.mnstate.edu/$\sim$juan/Research/HTDC/} for download.   Each catalog contains a star ID number, right ascension and declination (in mean J2000 coordinates), errors in the position, star/galaxy classification, the $V$ magnitudes and  $U-B$, $B-V$ and $V-R$ colors, and their associated uncertainties.  The 
format of the catalog is illustrated in Table~\ref{tbl3}.

\section{The  Star Count Analysis}

For the star count analysis discussed in this section, we have corrected the observed magnitudes 
and color in the catalogs   for interstellar extinction, determined the completeness limits of our 
fields, and calculated the corresponding coverage on the sky. 
Stars fainter than V = 16 mag
 down to the completeness limit of each field are included in the analysis.

\subsection{Interstellar Extinction}

Because interstellar extinction can be significant, especially for our fields below
{\it b} $=30\arcdeg$, all of our data is corrected for interstellar reddening using
the H I -- infrared dust emission maps and extinction tables from \citet{1998ApJ...500..525S}.  While 3-dimensional maps and other improvements are now available, the Schlegel et al. maps  have the sky coverage
required by our field placement and has generally been shown to be accurate for the higher
Galactic latitudes where we are observing.  For extremely faint red stars (e.g. those near the Sun) we only apply a fraction of the total correction using a Galactic model, described below,
and an assumed exponential dust distribution with a scale height of 100 pc.

\subsection{Completeness Limits}

The faintest objects in our fields vary from fainter than 20th to
 23rd magnitude in V and in the different colors; however
the completeness limit for our star counts will be significantly
brighter, especially for the restricted color ranges we use for the 
population separation.  In Paper I we showed that due to the disk
stellar density function, the classic technique of determining
completeness from the observed luminosity function, log N vs. apparent
$V$ magnitude will set the completeness limit too bright. In this paper, we have
conservatively adopted this classic estimate of completeness,
 effectively setting the completeness limit one magnitude
brighter than we otherwise might be able to claim if we used a
model-based interpretation.

To minimize questions about completeness limits and
model-based corrections,  we have tried not to push the statistics
of our faintest stars. 
Due to their relatively high galactic latitudes, confusion is minimal in
our fields.  Furthermore, the need for model interpretations
is minimized by  the symmetric field placement in Q1 vs. Q4.
Even for our worst case images, with the smaller telescope at CTIO, the SNR
for an V=19 star is greater than 40 and does not actually 
constitute a low signal to noise detection.  
We claim that within our stated completeness limits we are
better than $99\%$ complete and so no completeness corrections are required.

We are also fortunate that the intrinisic color dispersion in our fields does not appear to differ from field to field.  No corrections for differing amounts photometric scatter to normalize our color distributions have been required.

The faintest, reddest stars in our survey within the 
limiting V magnitude often lacked a measured $B-V$ color.
For these stars with $V-R$ colors only, a $B-V$ color was estimated using these
relations determined from fits to the Landolt standards \citep{1992AJ....104..340L}:

\[
B-V = 1.782(V-R) - 0.050,\;  V-R \leq 0.758
\]
and
\[
B-V = 0.802(V-R) + 0.689,\;  V-R > 0.758
\]
In practice most of the stars corrected in this way are much fainter than the stated completeness limit, and play only a small role in the analysis.

\subsection{Sky Coverage}

The CCD fields do not all have the same projected area on the sky. 
For each field in this program we have therefore measured the sky
coverage by analyzing the overlaps from the individual image WCS solutions.
Because these counts are always used in comparisons between paired fields we
normalize our fields by the relative areas of each field.  
To compute the normalized ratio $R$ between any two fields with areas 
$A_1$ and $A_2$ containing star counts $C_1$ and $C_2$, we use the formula:
 
\[
         R_{12} = \frac{C_1 A_2}{C_2 A_1}
\]

\subsection{Population Separation}

To identify the stellar population responsible for the asymmetry and  map
the spatial distribution of the Hercules Cloud,
we must discriminate among the major components of the Galaxy: the Disk, the
Thick Disk and the Halo. In Paper I, while searching for evidence of triaxiality,
we simply compared the number of faint
blue stars, counted bluewards of the ``blue ridgeline," the peak or maximum
of the color-magnitude diagram which occurs near a $B - V$ color of about
0.6 (hereafter any color representing the blue ridgeline will be denoted by the subscript ``P",
example $(B-V)_P$) over a large magnitude range. The stars identified this way will 
typically be
dominated by the Halo and Thick Disk stars,  but this technique does not
identify which population of stars is responsible for the excess.

Consequently, we use our three component star count model GALMOD
\citep{lar03} with
its best-fit parameters  to predict
the expected contributions of the Disk, Thick Disk and Halo stars to the
observed star
counts as a function of magnitude, color and direction.
Parameters determined in \cite{lar03} are safely within the wide
variation of
galaxy model parameters currently in the literature (see
\cite{2010ApJ...712..692C} for a discussion).  One exception is the radial scale length of the thick disk.  GALMOD's
value of 4500 pc is high compared to most other work
\citep{2003A&A...409..523R,2005A&A...436..895G,2006A&A...451..125V,jur08} but due to the large covariance that exists
between many galaxy model parameters, GALMODs 88 field fit
agrees to better than 3\%  with the Padova model
when predicting the fractional composition of stars by component and magnitude \citep{2006A&A...451..125V}. 

\begin{figure}
\epsscale{0.8}
\plotone{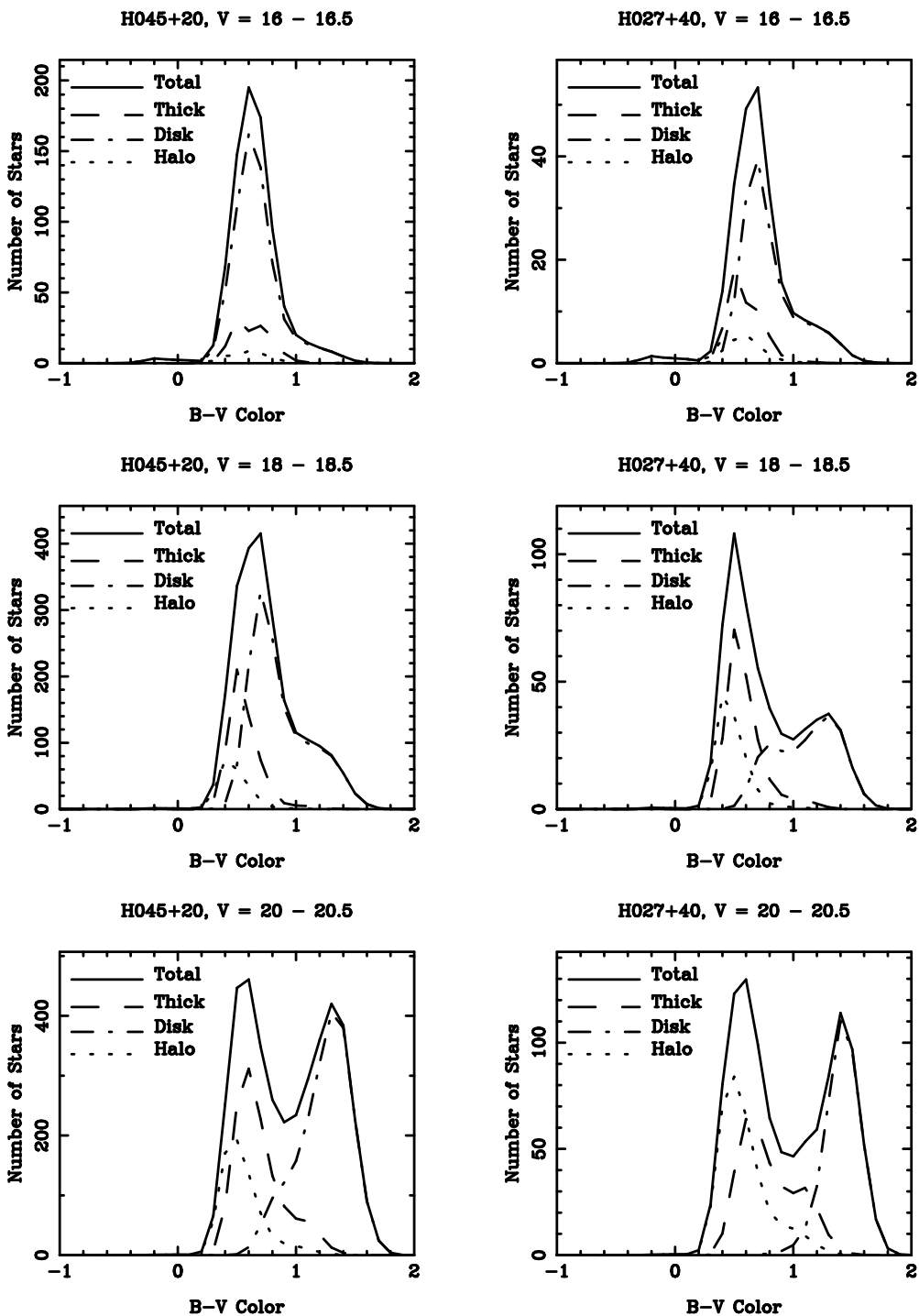}
\vspace{0.5in}
\caption{Example GALMOD output predicting the relative contributions of Disk, Halo and Thick Disk for two different directions ($l=45\degree$, $b=20\arcdeg$ and $l=27\degree$, $b=+40\degree$.) \label{fig2}}
\end{figure}

\begin{figure}
\epsscale{0.8}
\plotone{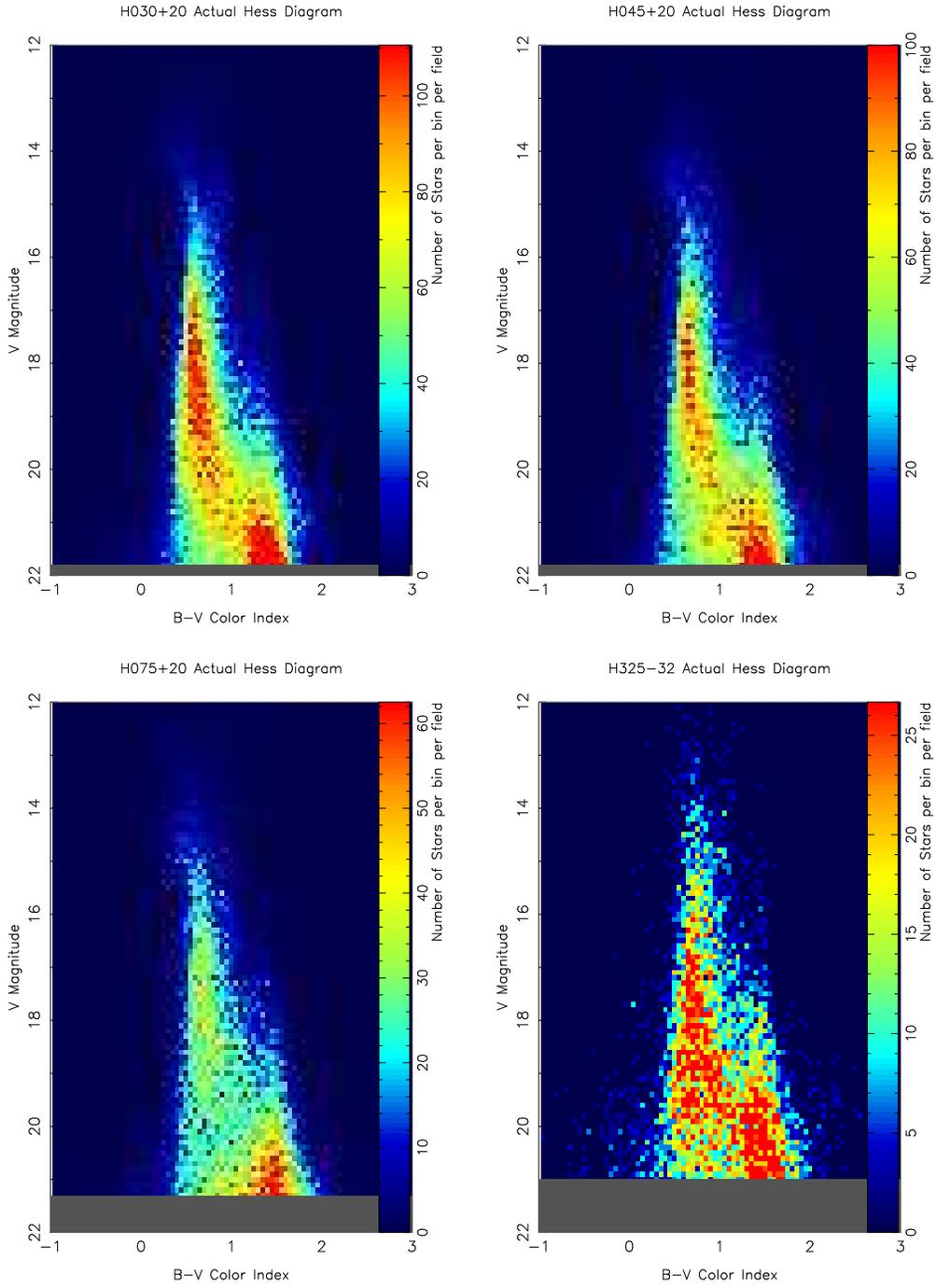}
\vspace{0.5in}
\caption{Hess diagrams from our star catalogs for four program fields (H030+20, H045+20, H075+20 and H325-32). \label{fig3}}
\end{figure}

\begin{figure}
\epsscale{0.8}
\plotone{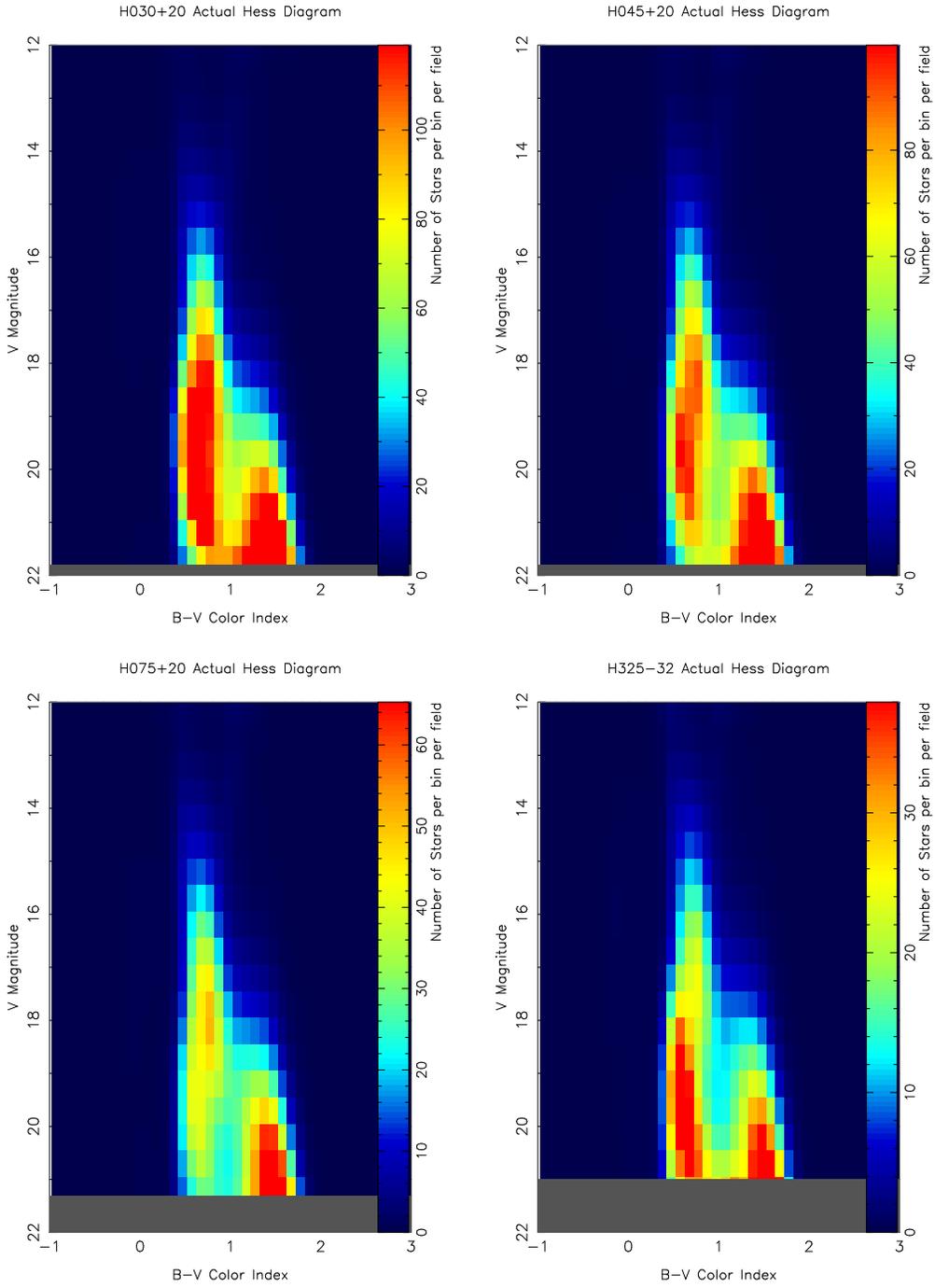}
\vspace{0.5in}
\caption{GALMOD generated Hess diagrams for the same four program fields as Figure~\ref{fig3}.  Bins have been scaled to compensate for their different sizes. \label{fig4}}
\end{figure}

Example outputs from GALMOD illusrating the contributions of
each component are in Figure~\ref{fig2}.  In addition, GALMOD is useful
for predicting ratios for fields  to compensate for differing lines of
sight or when the lines of sight penetrate a population (like the Disk)
in a
non-symmetric fashion.  As a demonstration of the fit parameter
validity, Figure~\ref{fig3} presents four sample Hess diagrams created
from our star catalogs (two-dimensional
 number histograms, binned by magnitude and  $B-V$ color).  This data
 can be readily compared to the GALMOD predictions for the same direction
 in Figure~\ref{fig4}.
The data and the
model show a qualitative agreement, which is sufficient for the
needs of our discussion.

In practice, the location of the blue ridgeline is influenced by several physical effects:
magnitude calibration zero points, the relative contributions of Disk, Halo and Thick Disk,
and possible variable uncorrected extinction.  To determine the location of the ridgeline
in our extinction-corrected color-magnitude diagrams, we initially estimate a location by eye
and then compute the median $B-V$ color from the stars selected within 0.4 dex.
The ridgeline locations are used later to compare the star counts in the same color 
ranges 
between our various fields.  Examination of color-number histograms
shows that the median ridgeline determination brought the color systems of
our various fields into agreement in more than 90\% of the cases.  For the
remainder an additional correction to the colors was required which
was never larger than 0.04 dex.  Because differences in the ridgeline 
location of 0.01 dex were very apparent in color-number histograms we conclude
that the remaining error in color zero points between fields is no larger than this.

Table~\ref{tbl4} gives the sky coverage, completeness, and ridgeline location in the magnitude ranges $16 < V < 18$, $18 < V < 19$, $19 < V < 20$ and $20 < V < 21$ for all of the
program fields.

With the peak of the blue ridgeline as a reference point, we then define three
population or color groups. For comparison with \cite{par03} we adopted color ranges in $B-V$ corresponding to the MAPS photometric system  \cite{1996PhDT........83L}. 
The color ranges used in this paper 
with respect to the (B - V$)_{P}$ are: ``Blue" ( $-1.0 < B-V < (B-V)_P $ ),
``Intermediate" ( $(B-V)_P - 0.05 < (B-V) < (B-V)_P + 0.5 $ ) and
``Red" ( $(B-V)_P + 0.5 < B-V < (B-V)_P + 1.5$ ).

\subsection{The Star Count Ratios -- Definitions}

For this analysis, we determine the star count ratios for paired fields across the $l=0\degree$ line between
Q1 and Q4 and also across the Galactic plane ($b=0\degree$) within Q1 or Q4.
Each ratio is computed in the order of the comparison with the numerator being the first field and the denominator being the second field.
For each pair, we compute the ratios in five magnitude ranges:  $16 < V < Completeness$, $16 < V < 18$, $18 < V < 19$, $19 < V < 20$ and $20 < V < 21$ for each color range.  The joint completeness limit for a matched pair is
set by the field with the brighter limit.  

GALMOD was first used to predict the net count ratios and fraction of stars belonging to each stellar component (Disk, Halo, Thick Disk) for all of the color and magnitude ranges where the field pairs are complete.  The results  are presented in Appendix A, Tables~\ref{tbl5} (``Blue"), \ref{tbl6} (``Intermediate") and \ref{tbl7} (``Red").
In general, the model-based ratios  are very close to unity if we compare across the $l=0\degree$ line of symmetry and is less than one if we compare across the $b=0\degree$ line of symmetry due to the Sun's position above the Galactic disk.  The fractions of stars due to each stellar component show a general trend when divided into our color ranges.  For the ``Blue" color bin (Table ~\ref{tbl5}) the Disk is only represented strongly for bright magnitudes and low latitude fields.  As the magnitudes move fainter or the line of sight to higher latitude the Thick Disk and Halo dominate.  In the ``Intermediate" range (Table ~\ref{tbl6}) the Disk still dominates the brighter magnitudes and low latitudes but the Thick Disk becomes preferred to the Halo in the fainter and higher latitude fields.  For the ``Red" color cut (Table ~\ref{tbl7}) the Disk strongly dominates all magnitudes and lines of sight.

We then computed the observed star count ratios ($R$) for the same magnitude and color range used in the
model computations with the caveat that the number of stars are normalized to the same area. 
The uncertainties in the number of stars in each field ($N$) are computed  considering
two main effects;  Poisson error in the number of counts ($\sigma_N$), and the effect of
an error in the color zero point  on the measured number of stars ($\sigma_P$).  The second error was determined by changing the color limits by the allowed zero point error and seeing how the number of stars would change.  The net error in the number of stars in one of our fields $i$ is then given by adding the two possible errors in quadrature:

\[
 \sigma_i = sqrt{(\sigma_N^2 + \sigma_P^2)}
\]

The uncertainty in the ratio $R$ between two fields with number of stars $N_1$ and $N_2$ can be calculated from:

\[
 \sigma_R = \frac{N_1}{N_2} \sqrt{(\frac{\sigma_1}{N_1})^2 + (\frac{\sigma_2}{N_2})^2} 
\]

Finally, the interpretation of the significance of a ratio's difference from the GALMOD prediction is computed following \citet{par03} by first defining a ``super-ratio" between the ratio ($R$) and the GALMOD prediction ($R_{GALMOD}$) for the ratio and then examining it's significance ($s$) compared to unity given the uncertainty on the ratio:

\[
  R_S = \frac{R}{R_{GALMOD}}
\]

\[
 s = \frac{(R_S - 1)}{\sigma_R}
\]

Our significance values tend to be much smaller than in \citet{par03} because our higher photometric precision is compromised by the smaller areas and 
therefore fewer stars.

The star count ratios from our catalogs, combined errors, and significance parameters for the
``Blue," ``Intermediate" and ``Red" color ranges in Appendix B, in Tables~\ref{tbl8},
\ref{tbl9} and \ref{tbl10}, respectively.

\subsection{The Star Count Ratios -- Results}

Examination of Table~\ref{tbl8} (``Blue") shows a large number of moderately significant ratios $>$ 1 for comparisons across the $l=0\degree$ line for the magnitude ranges $16 < V < 18$.  Most of these ratios are above the plane but three are below.  This excess rapidly fades and is mostly gone before 19th magnitude in $V$.  By $19 < V < 20$ there are a handful of fields showing a deficit of blue stars with respect to Q4 at moderate $l$ and $b$ of similar significance.

The ``Intermediate" ranges weakly echo the behavior of the blue ranges in that the same fields in Q1 and above the plane show the excess but in general both the ratio and its significance parameter are smaller.  The same directions show the mentioned deficit of stars between $18 < V < 19$.  

In the ``Red" ranges, a large number of Q1 fields show a red star excess mirroring the ``Blue" excess in the same location.  This excess generally exists a magnitude fainter, but at colors different enough that the stars exist in other regions of the Galaxy.  The above/below the plane excess at fainter magnitudes may represent an incorrect Disk scale height in the GALMOD fit.

\begin{figure}
\epsscale{0.8}
\plotone{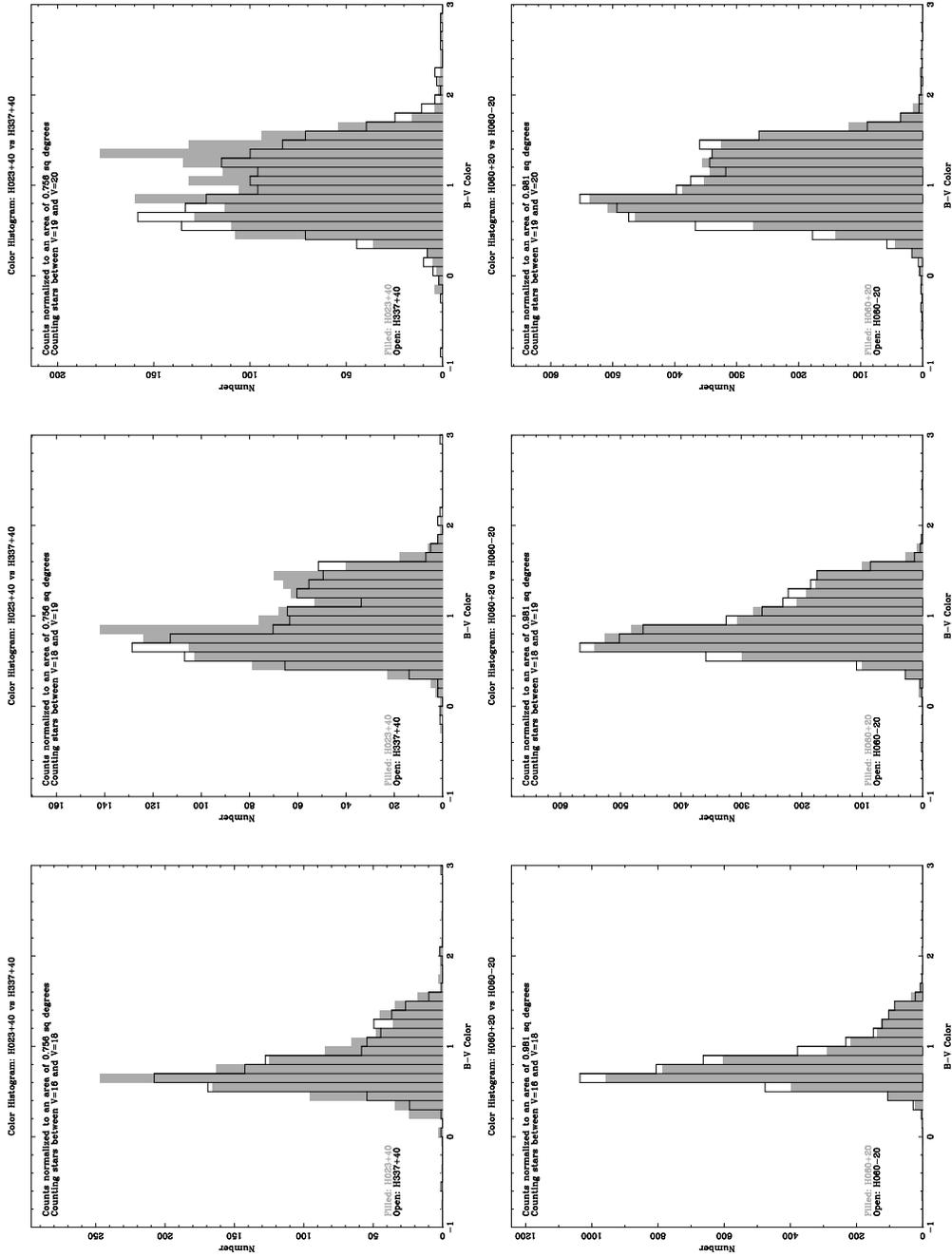}
\vspace{0.5in}
\caption{Color histograms for a field which shows the star count excess (H023+40/H337+40, top three panels in order of increating magnitude) and a field which does not show an excess (H060+20/H060-20, bottom three panels in order of increasing magnitude). In the bottom three panels a small preference for the below-the-plane field is expected because of the Sun's position relative to the Galactic midplane.\label{fig5}}
\end{figure}

The star count excesses can be easily displayed.  In Figure~\ref{fig5}, the top three panels display color histograms over different magnitude ranges.  Excesses in both ``Blue" and ``Red" color ranges are apparent.  Note how the blue excess fades as the magnitude gets fainter while the red excess continues to grow.  The bottom three panels in the same figure illustrate a field which does not display a significant excess.

To assist in the visualization of the ratios we show the deviation of the data from the smooth model predictions, as the super-ratio, along with its significance as a function of position on the sky.  In  Figures 6 and 7 , we plot the program fields (altered in some cases to prevent overlap) and shade the field by the significance parameter.  Over this, a symbol is placed expressing the net size of the deviation of the data from the model.

An examination of the super-ratios for all ``Blue" stars across the $b=0
\degree$ line of symmetry from V=16 to the completeness limit for the combined fields shows that, apart from a general agreement with the model predictions, the largest
discrepancy is our highest latitude field.  No significant deviations
are found across this line of symmetry.

\begin{figure}
\epsscale{0.6}
\plotone{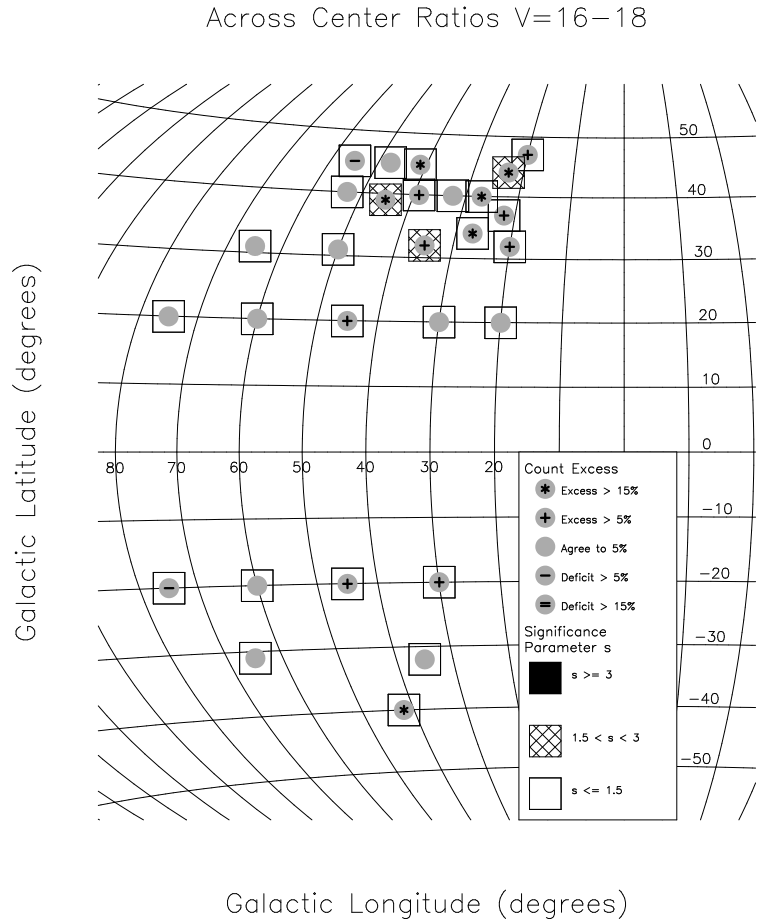}
\epsscale{1.0}
\vspace{0.5in}
\caption{``Blue" super-ratios with significance parameter across $l=0$ with $16 < V < 18$. \label{fig7}}
\end{figure}

In Figure~\ref{fig7} we see a wide-ranging asymmetry in the faint blue
stars counts in Q1 over Q4.  The asymmetry fades with fainter magnitudes
and essentially vanishes by $V=20$.  Curiously, an examination of 
Table~\ref{tbl9} shows that there is a range of intermediate colors ($0.6 \leq B-V \leq 1$)
where
the excess does not appear even though it is present in both ``Blue" and
``Red" (see Figure~\ref{fig5}).
In any case, the ``Red" color range displays the excess again, extending
to fainter magnitudes as can be seen in Table~\ref{tbl10}.

\begin{figure}
\epsscale{0.6}
\plotone{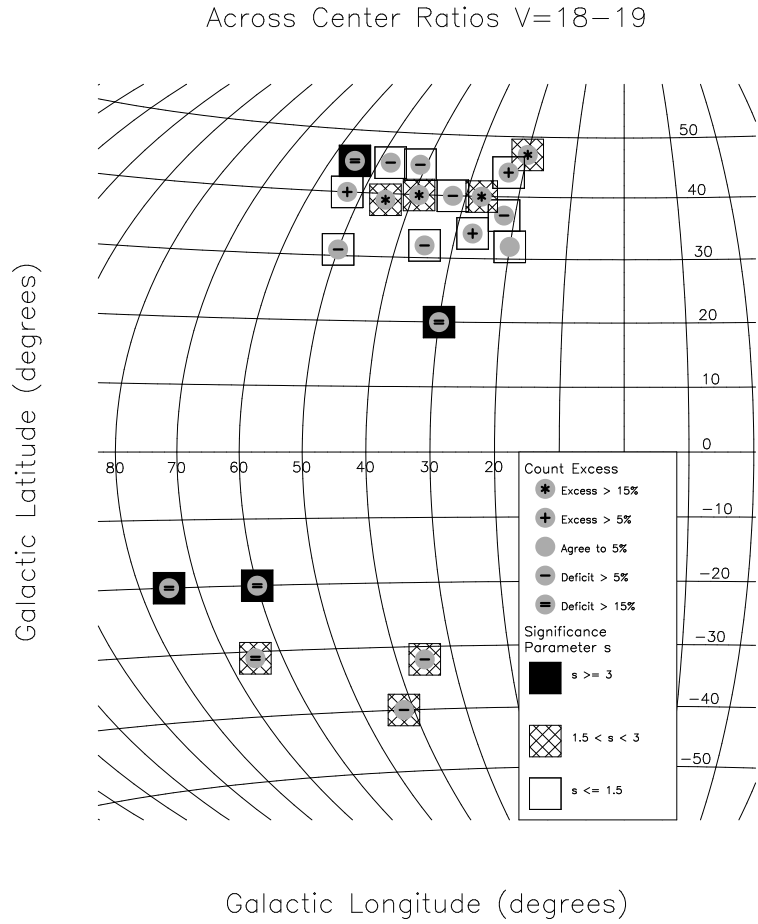}
\epsscale{1.0}
\vspace{0.5in}
\caption{``Red" super-ratio with significance parameter across $l=0$ with $18 < V < 19$. \label{fig9}}
\end{figure}

A surprisingly  large excess of red stars occurs in
Q4 below the plane.  Figure~\ref{fig9} shows ``Red" stars compared
across $b=0\degree$.  In Q4, many fields show a significant
deficit above the plane compared to the corresponding fields below the
plane.  Examination of Table~\ref{tbl10} also demonstrates that the excess is
IN Q4 below the plane because the  Q1 stars below the plane are outnumbered
by their counterparts in Q4.  We merely comment on this observation.
It is outside the scope of our interest in the faint blue stars above
the plane and intend to follow up on this interesting observation in a
future work.

\subsection{Identification of the Hercules Thick Disk Cloud with the Thick Disk}

It is difficult to separate the Halo and Thick Disk in our data at our relatively bright limiting magnitudes, using only color.  Most of this problem is due to our relatively small field size compared to Parker and due to our decreased color dispersion which does not create an extreme blue tail of Halo stars.  A simple
correlation analysis, however, 
suggests that the Hercules Cloud is identified most strongly with the Thick Disk.

\begin{figure}
\epsscale{0.5}
\plotone{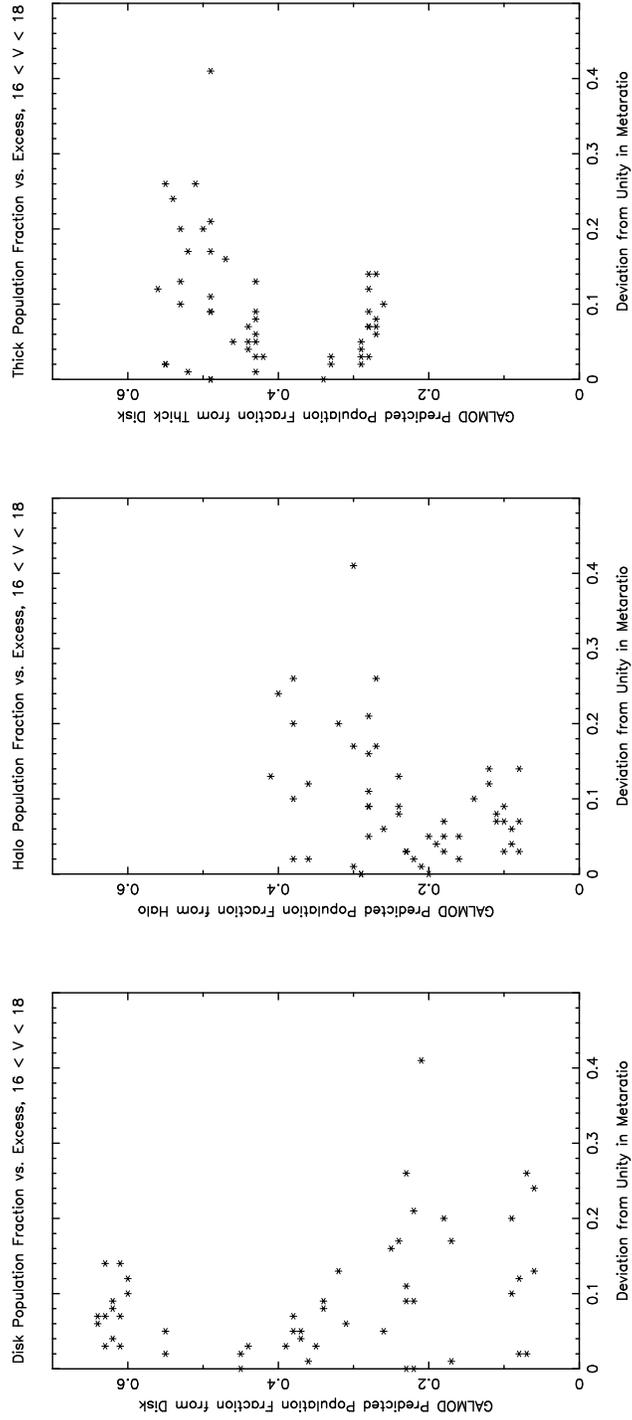}
\vspace{0.5in}
\caption{GALMOD predicted strength of excess by fraction of population in each component for $16 < V < 18$. \label{fig11}}
\end{figure}

Figure~\ref{fig11} shows the GALMOD predicted fraction of stars for the Disk, Halo and Thick Disk vs. the deviation of the super-ratio from unity for ``Blue" stars with magnitudes $16 < V < 18$.  For the Disk (left panel), the deviations occur most strongly when the Disk star fraction is low.  Since the excess occurs both above and below the plane, we infer that a large number of Disk stars
in similar magnitude and color ranges may be hiding the signature of the excess in the low latitude fields.  For the Halo plot (middle panel), the higher deviations are strongest with higher Halo star fractions but there is a fair amount of scatter in this correlation.  For the Thick Disk, however, the right panel of Figure~\ref{fig11} shows that higher star fractions correlate with higher deviations like the Halo. Unlike the Halo, this trend seems to possess
a much tighter correlation.
Based on this set of plots we suggest that at the lower latitudes the Disk dominates the star counts and dilutes the strength of the excess to the level where we would need much larger areas to detect it statistically.  This would also explain why the excess 
is strongest at higher latitudes, where the lines of sight do not contain so many Disk stars.

We also note the excess does not appear at latitudes higher than $b \sim 40\degree$.  This is consistent with a density distribution  which our line of sight exits at some height which would not be the case even with a flattened Halo density function.  Choosing between the Halo and the Thick Disk, the excess appears more strongly identified with the Thick Disk.  Because we see the ``Blue" ratios in Table~\ref{tbl8} decrease back towards unity for fainter magnitudes, we believe that we have seen through the HTDC in these directions.

\section{Location of the Hercules Thick Disk Cloud within the Galaxy}

Our primary goal with this study is to map the over-density or star count excess
and determine its spatial extent along the line of sight. In the previous section 
we have confirmed the excess among faint blue stars in Q1 above and below the plane,
and identified it with the Thick Disk population. 
In this section we use the method of photometric parallaxes to estimate the distances
and map the star count excess along several lines of sight in Q1 to determine the 
size and mass of the Hercules Cloud. 

\subsection{Photometric Parallax}

\begin{figure}
\epsscale{0.7}
\plotone{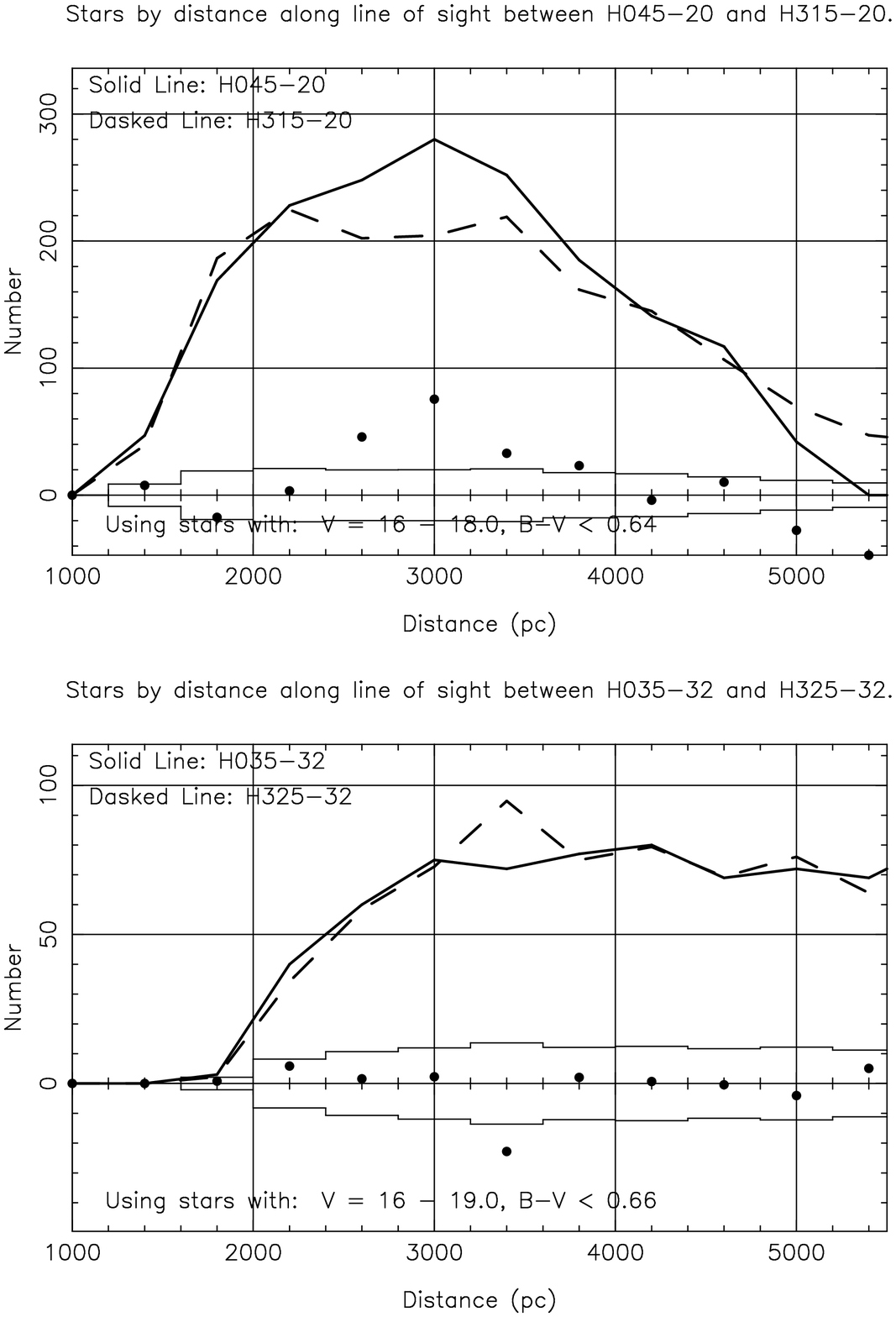}
\vspace{0.5in}
\caption{Number vs. distance for two paired lines of sight determined through photometric parallax.  Both plots include stars in the defined completeness range and colors have been selected to be consistent with the ``Blue" color selection.  The top plot shows the parallax for a field displaying the excess, the bottom for a field without the excess.  On each plot the solid line is the number vs. distance for the first field in the ratio, the dashed line is for the second field in the ratio, the dots represent the difference between the two lines and the histograms represent the poisson error on the difference. \label{fig12}}
\end{figure}

\begin{figure}
\epsscale{0.7}
\plotone{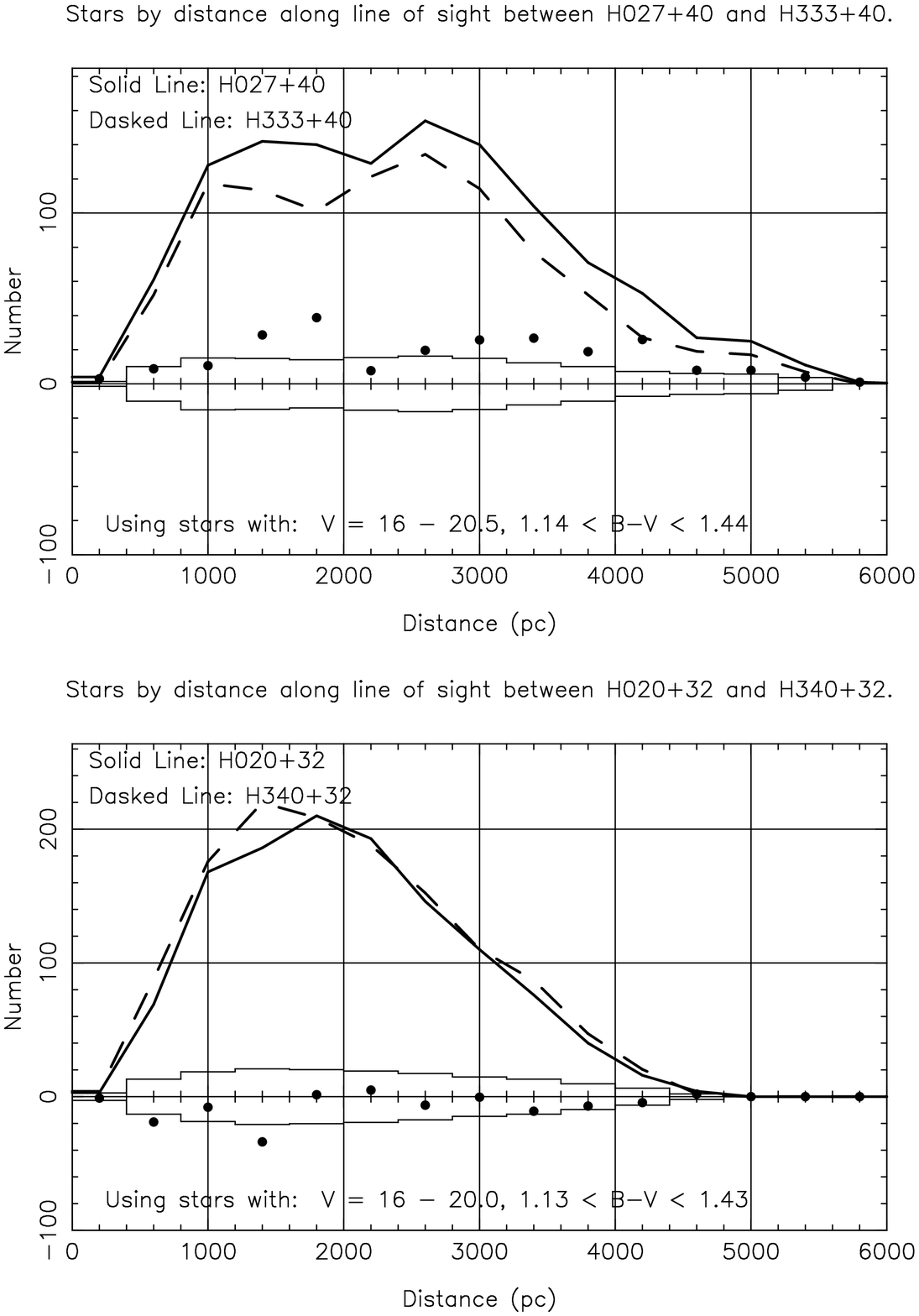}
\vspace{0.5in}
\caption{Number vs. distance for two paired lines of sight determined through photometric parallax.  Both plots include stars in the defined completeness range and colors have been selected to be consistent with the ``Red" color selection.  The top plot shows the parallax for a field displaying the excess, the bottom for a field without the excess.   On each plot the solid line is the number vs. distance for the first field in the ratio, the dashed line is for the second field in the ratio, the dots represent the difference between the two lines and the histograms represent the poisson error on the difference.\label{fig13}}
\end{figure}

Our count ratio analysis shows that the asymmetry and star count excess is associated 
most strongly with the ``Blue" and ``Red" color ranges.  We use the method of 
photometric parallaxes to derive typical distances for these stars.
To  proceed,
we adopted the Thick Disk color-magnitude diagram and relative luminosity function  from
\cite{1983MNRAS.202.1025G}, used in GALMOD.
We select stars  in our  ``Blue'' and ``Red'' color bins with magnitudes  between $V = 16$
and the  completeness limit  in the paired fields across the $l=0\degree$ line of symmetry.     The application of photometric parallax is straightforward.  Using the extinction-corrected
 magnitudes and colors,  the photometric distance is computed from the adopted Thick Disk
 color-magnitude diagram and luminosity function.  Figures~\ref{fig12} (``Blue") and \ref{fig13} (``Red") show the number of stars in 400 pc wide bins vs. distance along two paired sample
 lines of sight.  It is clear from these examples that although we have detected an excess 
integrated along the line of sight there is variation in the ratio with distance.  Since we have taken a detection of moderate statistical significance and separated it into many smaller distance dependent bins, the statistical
significance of each of these bins is less. We expect this rebinning to be
noisy, and as a result we use these results only to map the location
of the star count excesss in Galactic coordinates. We therefore restrict our distance ranges at each longitude and latitude to those
 where the star count in Q1 exceeds Q4 by more than one sigma.
Considering the tightness in the color range, which occurs within 0.15 dex of the Thick Disk turnoff color, photometric uncertainty could be responsible for up to 50\% errors on the inferred distances of some individual stars.  

\begin{figure}
\epsscale{1.0}
\plottwo{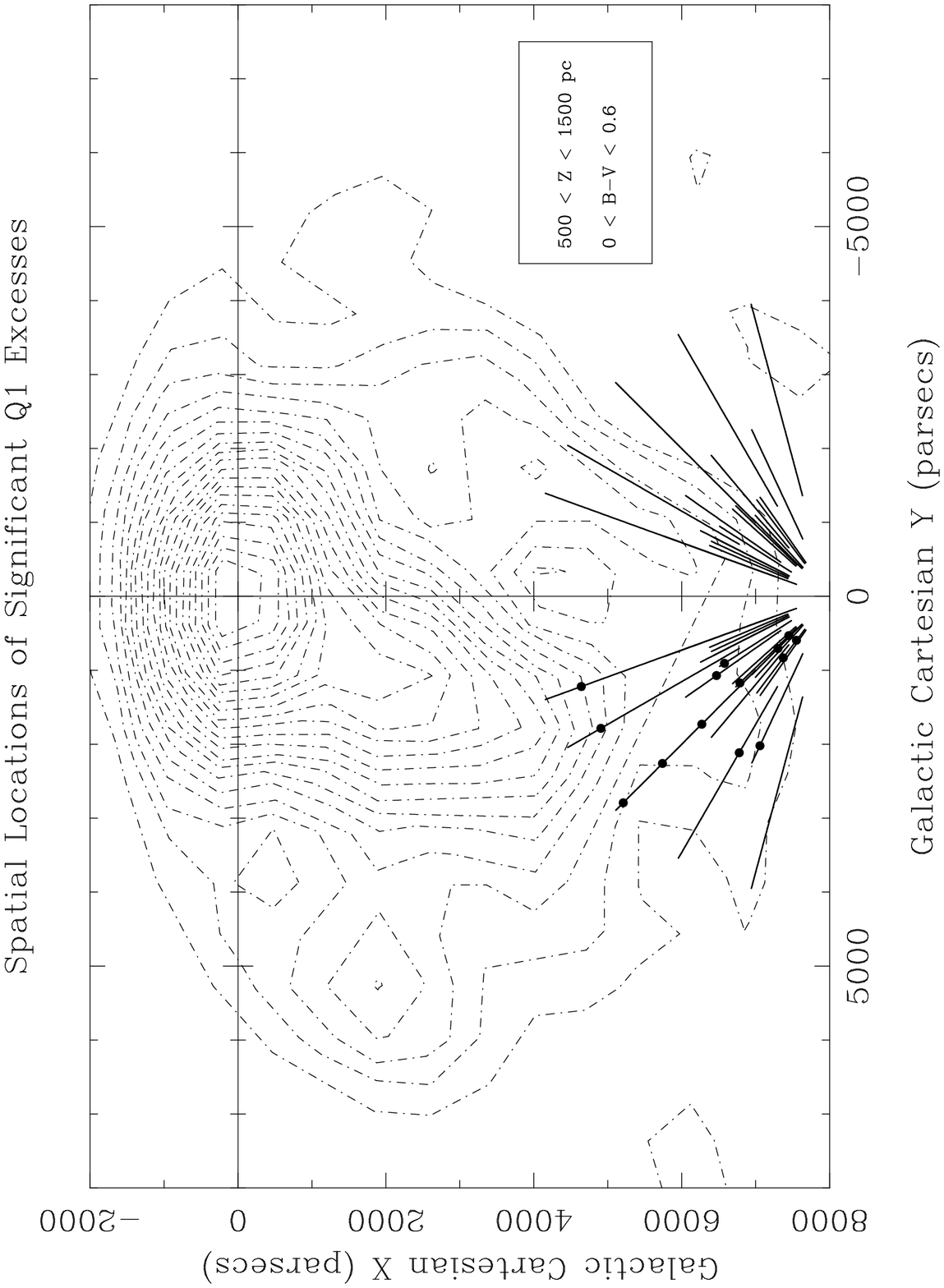}{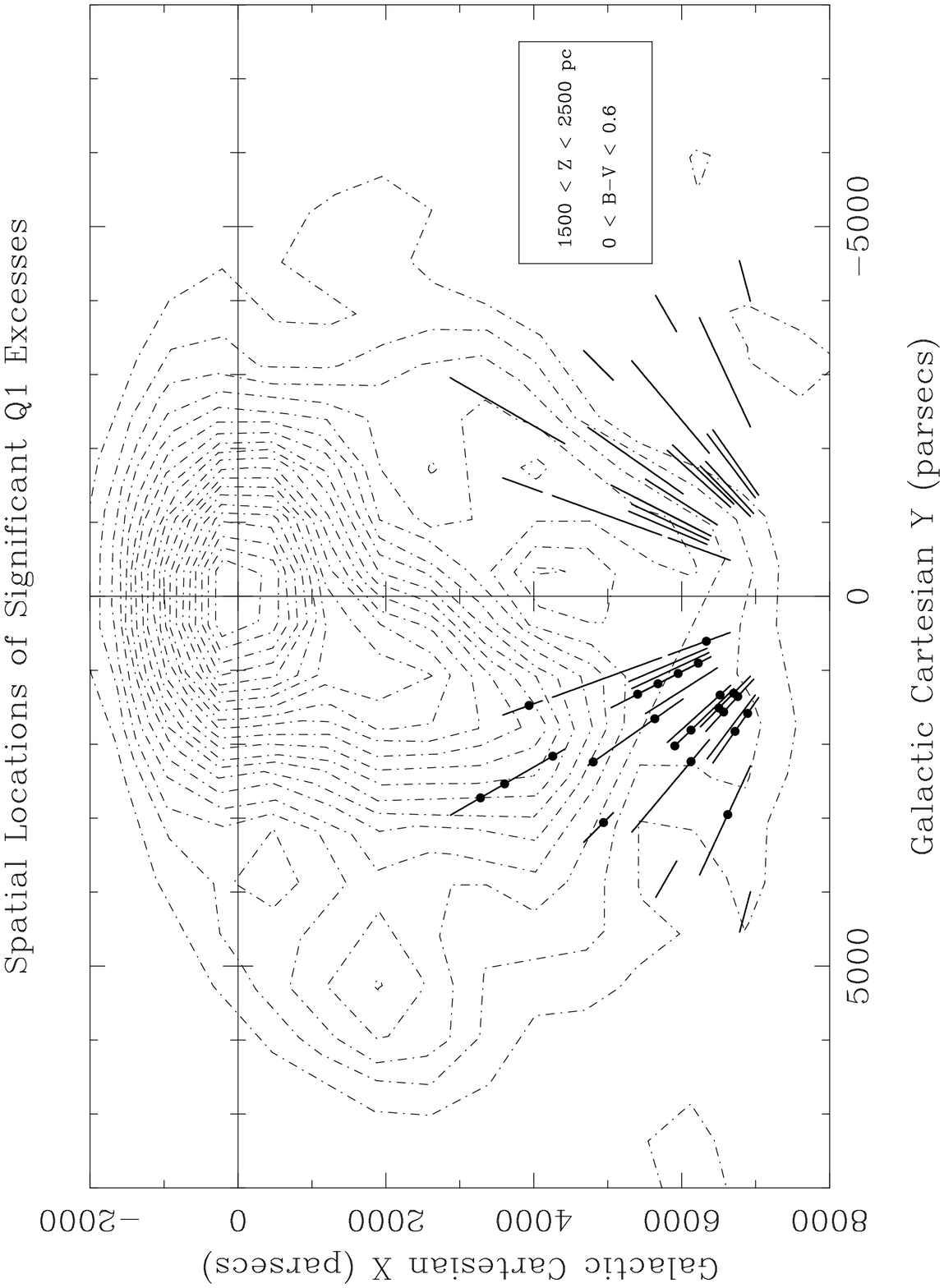}
\epsscale{0.47}
\plotone{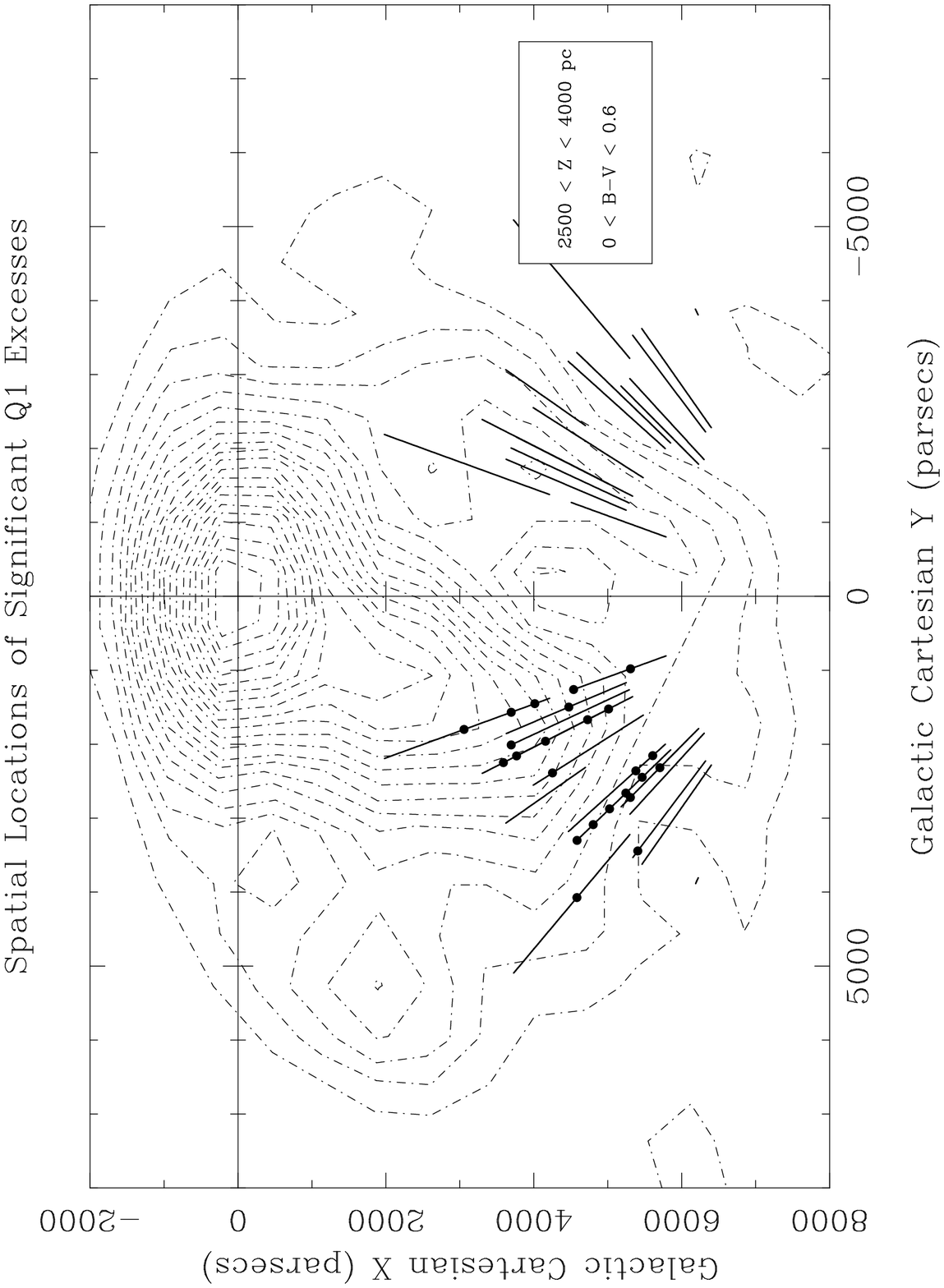}
\vspace{0.5in}
\caption{Location of excess for faint ``Blue" stars with (a) $500 pc < Z < 1500 pc$, (b) $1500 pc < Z < 2500 pc$ and (c) $2500 pc < Z < 4000 pc$ in galactocentric Cartesian coordinates. \label{fig14}}
\end{figure}

\begin{figure}
\epsscale{1.0}
\plottwo{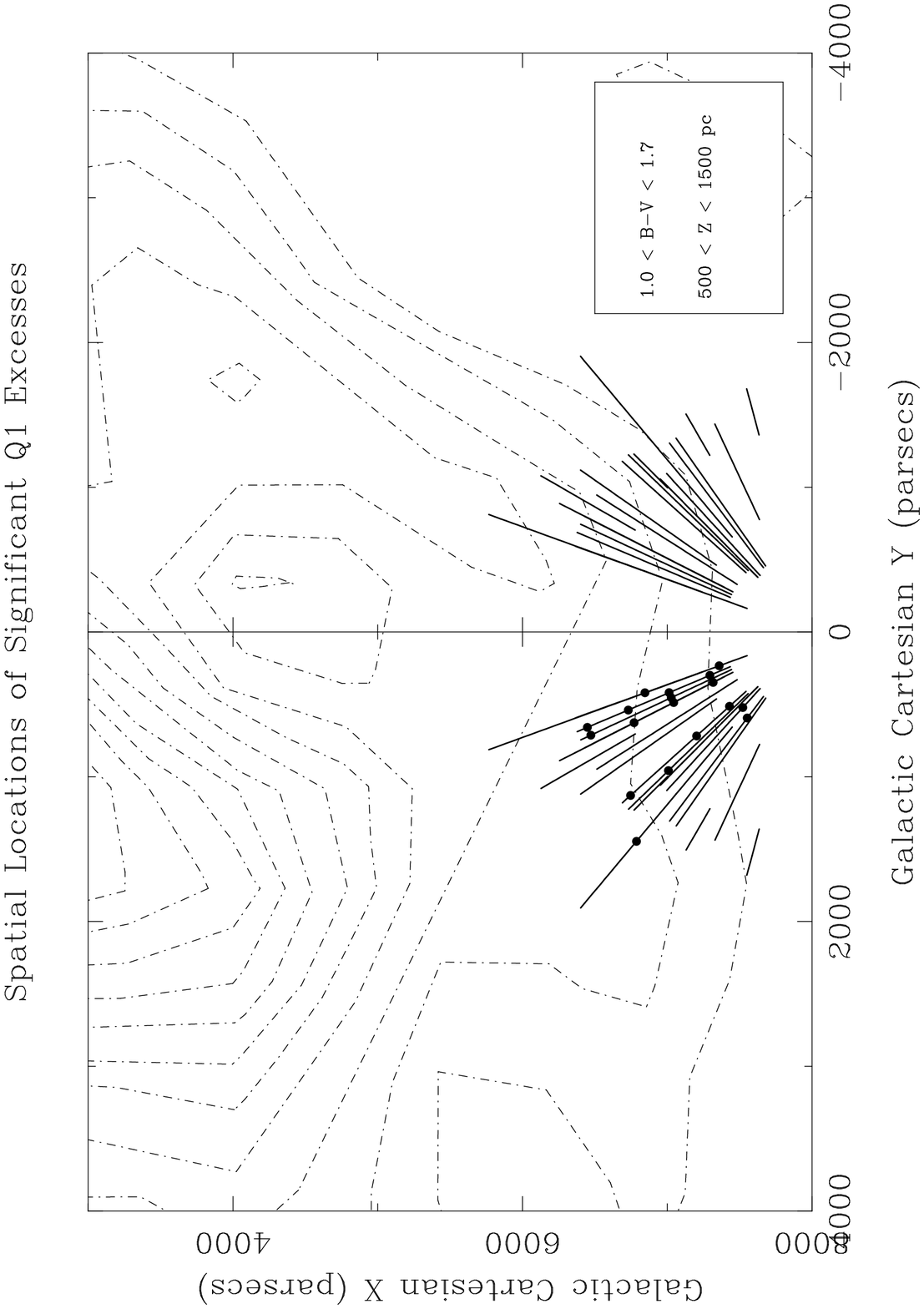}{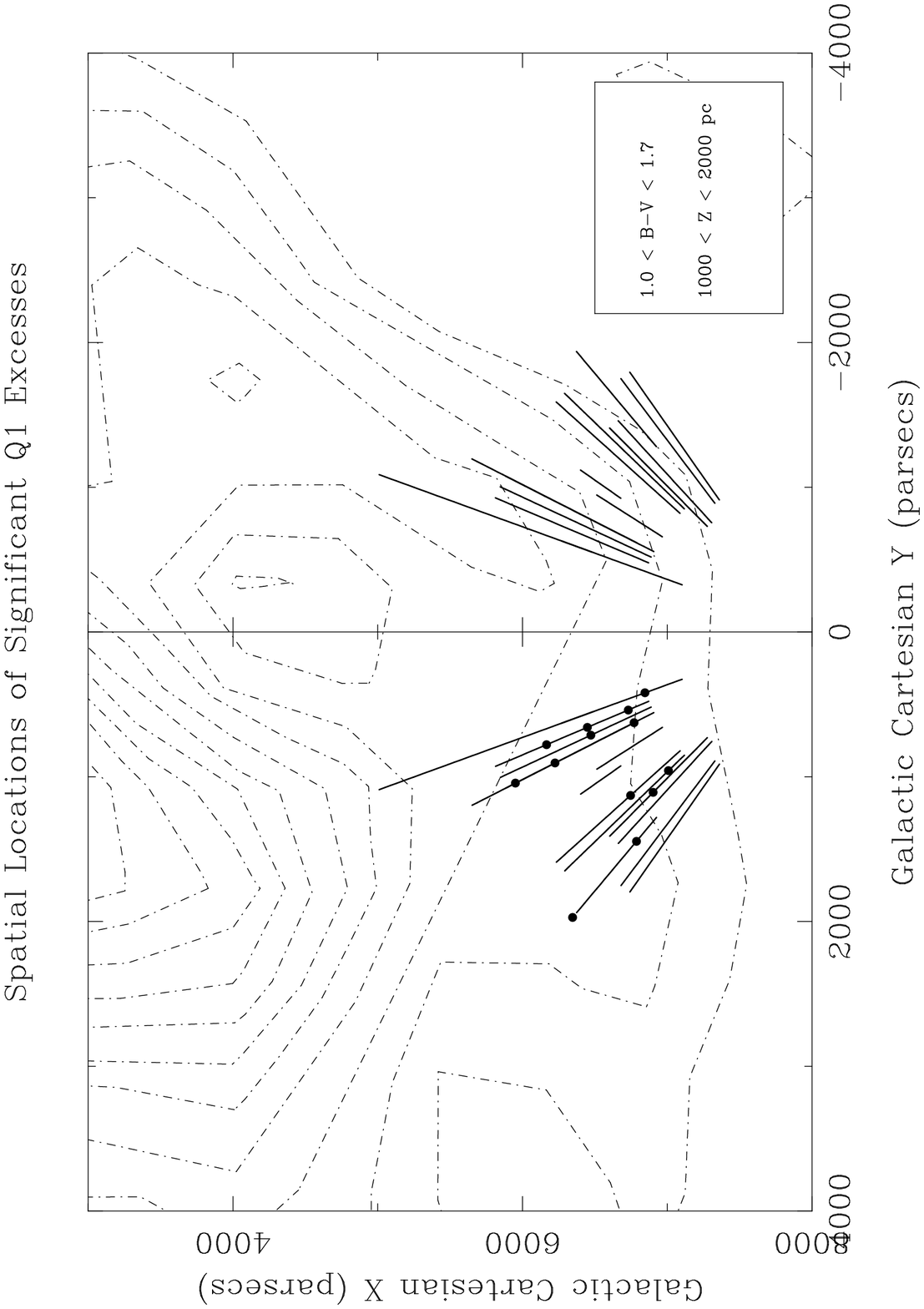}
\epsscale{0.47}
\plotone{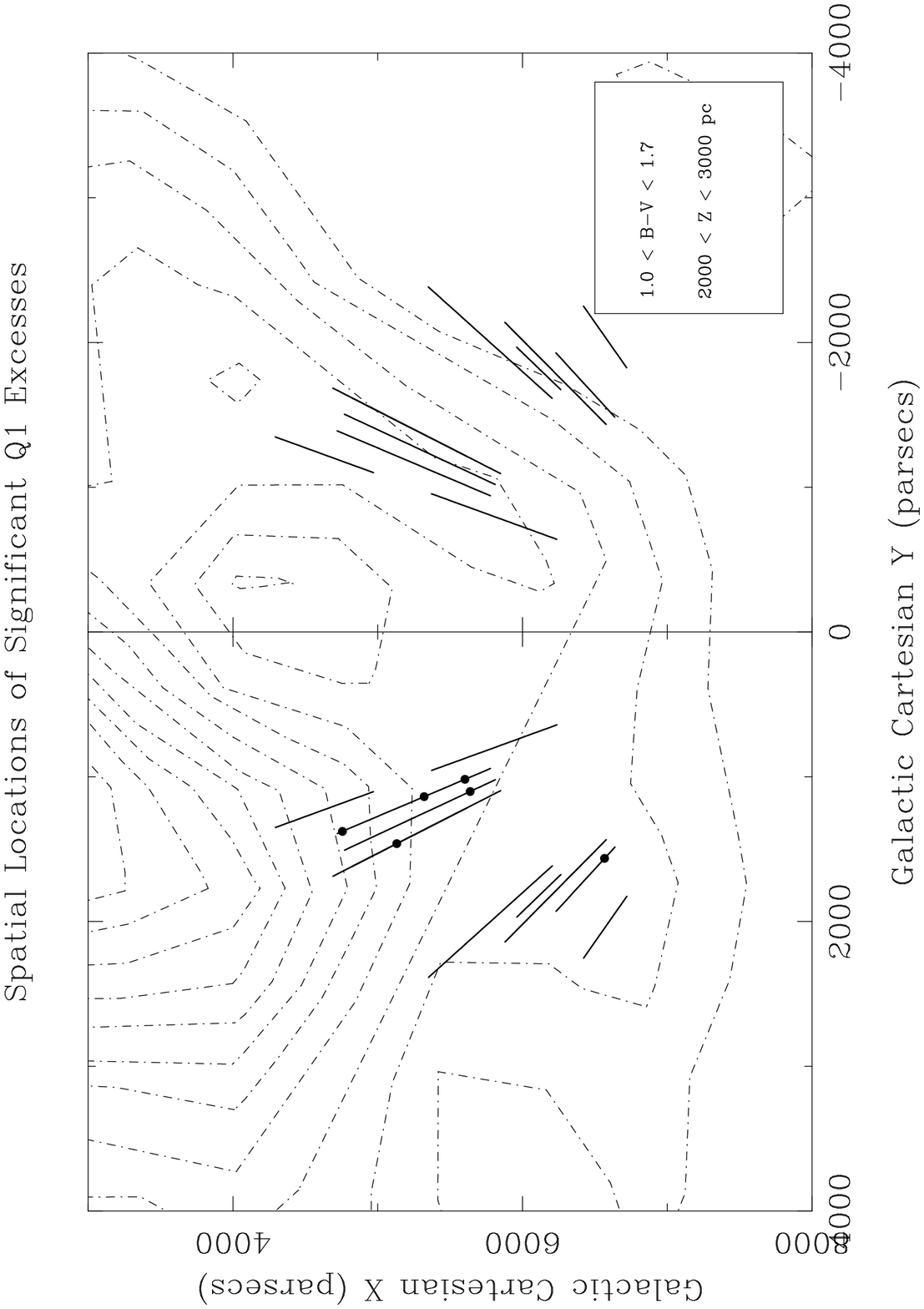}
\vspace{0.5in}
\caption{Location of excess for ``Red" stars with (a) $500 pc < Z < 1000 pc$, (b) $1000 pc < Z < 200 pc$ and (c) $2000 pc < Z < 3000 pc$ in galactocentric Cartesian coordiantes. \label{fig15}}
\end{figure}

We present our photometric parallax results in Figures ~\ref{fig14} and ~\ref{fig15}.  The plots are on a
Galactocentric Cartesian X,Y,Z coordinate system binned by distance $|Z|$ from the
Galactic plane.   All  of the figures  also show the density contours of  the bar in the
Disk as traced by IRAS AGB stars from \cite{1992ApJ...384...81W}.  The AGB stars are
confined to within 1 to 2 degrees of the Galactic plane and are therefore below  our
perspective in each figure.  The lines of sight to our program fields with $|Z|$  in the
indicated ranges are shown for both Q1 and Q4.   Note that many of the lines of sight
overplot each other due to identical $l$ and similar $b$.  To illustrate where the excess is not detected, we show only those sight lines where the catalogs would be complete in magnitude and color.  The dots on the Q1 lines of sight represent distances where the excess in Q1 exceeds the Q4 line of sight by more than one sigma.

The over-density regions for the ``Blue'' population
are shown in Figure
~\ref{fig14}  for $|Z|$ distances of  0.5-1.5 kpc,  1.5-2.5 kpc and  2.5 - 4.0 kpc.  Interestingly, the star count excess is strongest where our lines of sight cross the density contours 
of the bar and are associated with directions  where the bar's density appears to increase.
 At the lower $|Z|$ distances, the over-density regions are  not directly over the bar, 
but on the near-side. As we look at larger $|Z|$ distances, our line of sight also reaches larger 
distances, and the excess is more nearly directly over the bar. 
These lines of sight continue to even greater $|Z|$ than what are shown in
Figure ~\ref{fig14} and even fainter magnitudes.  At the greatest 
distances the excess disappears, since the ratios in Table \ref{tbl9} also 
return to close to unity; given that many of the fields were complete
to even greater distances our line of sight has apparently exited the 
cloud or the region showing the over-density.  

Figure ~\ref{fig15} shows similar illustrations for stars in the ``Red" color range with $|Z|$ 
between 0.5-1.5 kpc,   1.0-2.0 kpc and 2.0 - 3.0 kpc. Although the red star population is
closer, it shows the same trend as the ``Blue" group. The red star counts increase in the same 
longitude ranges and in the direction of  the bar.  
Although there are few data points at the highest $|Z|$ distances in Figure~\ref{fig15}, 
the star count excess  overlaps with the outer or nearer density contours of the bar.

\subsection{Size and Mass of the Cloud}

We have 63 lines of sight for the ``Blue'' stars, of which 14 in Q1 show
        an
        excess in the magnitude range $16 < V < 18$ compared with the
        corresponding Q4 fields.  This is not sufficient to clearly define the
        boundaries of the excess. The following discussion of the mass
        estimate is hindered by this  uncertainty.   Considering the results of \S
        4.1,
        we  use our ``Blue" stars as a tracer of the luminosity function
        and calculate the mass with the following  three assumptions for the extent of the
        cloud:
        
        \begin{itemize}
        \item A baseline assumption that the excess only exists in the 14 lines
        of sight and nowhere else.
        \item The excess is contained in the smallest volume which could
        geometrically enclose these 14 lines of sight.
        \item The excess is associated with some larger feature like the
        bar.
        \end{itemize}
        
        We assume that the Thick Disk luminosity function
        and color-magnitude relation used for the photometric parallaxes in
        \S4.1
        is valid.
        With this assumption, the colors of stars blueward of $B-V = 0.6$
        imply a range of absolute magnitudes in the Thick Disk luminosity
        function.
        Knowing how many stars with these colors actually belong to the excess
        in
        our 14 lines of sight, we can then ``normalize" the luminosity function
        and
        estimate the total number of thick disk stars the  excess  should
        represent.  This number is tied to the volume of space sampled.
        We can then scale the  result to any larger spatial volume. .
        While we have no information which lets us understand the radial
        number
        density distribution for the excess, we are looking at a high latitude
        population both above and below the disk.  If the over-density is
        gravitationally induced,   the star density of the excess will be higher near
	the plane of the
        galaxy and decrease with increasing $|Z|$.
        As an additional consideration, we adopt the expected exponential
        decrease in spatial number density for the Thick Disk  with a 
        $|Z|$ vertical scale height of approximately 900 parsecs.
        
        For our baseline assumption, we have 14 lines of sight covering 12.0 square degrees
        with 1,142 faint blue stars comprising the excess.  Given their  magnitudes
        and colors, these
        faint blue stars correspond to a total mass for the over-density of
        approximately 8,500 $M_\odot$ in just our lines of sight.  In \S4.1 we determined that the most
        distant
        stars in the excess are on average slightly more than 4.5 kpc from the
        Sun.
        Given the area of each field, our 14 lines of sight directly sample a
        volume
        of approximately 1.1\e{8} cubic parsecs and give a mass density for the excess of 8\e{-5}
        $M_\odot$/pc$^3$.
        
        For the second case, simple geometry shows 
        the minimum cube which would completely enclose
        all 14 of our lines of sight to their maximum distance 
        is 2.5\e{10} cubic parsecs.
        Assuming the density of our excess population is constant throughout,
        this volume would contain roughly 6\e{6} stars and has a total mass of
        some
        2\e{6} $M_\odot$.   If we assume that we are looking at a
        population whose vertical density scales as an exponential with the
        900 pc scale height we ascribe to the thick disk,  our detection would
        represent
        a higher associated mass of 2\e{7} $M_\odot$.
        
        Finally, if the excess is truly associated with the bar,
        the dimensions associated with the outer contours of Weinberg's bar in
        Figure ~\ref{fig14} and the height of our detection of the excess above the plane would
        imply a total volume of 6.0\e{11} cubic parsecs.
        At a constant number density, the Hercules Cloud would comprise some 1\e{8} stars
        with a total mass of 5\e{7} $M_\odot$.  Allowing for the vertical number
        density to scale as an exponential with a 900 pc scale height would
        increase these values to 1\e{9} stars and a total mass of 5\e{8} $M_
        \odot$, which then could
        comprise a
        substantial fraction of the presumed total Thick Disk mass.  This of
        course, would also have the implication that the Thick Disk
        is preferentially  aligned with the bar.
        For comparison, the most recent estimate for the mass of the Sagittarius
        Dwarf is 1.5-3.8\e{8} $M_\odot$  \citep{2010ApJ...714..229L}.
        
        The Gilmore and Reid luminosity function was chosen because of its long
        history.
        Newer luminosity functions \citep{2001A&A...373..886R}
        imply a different power law index for lower mass stars and therefore a
        smaller spatial number density, and would  lower our mass estimates by
        approximately 25\%.
        
        In summary it is highly likely that the Hercules Cloud is quite large.
        It's total mass may  range from a lower limit of
        2\e{6} $M_\odot$ if it is a local overdensity 
        with a potential upper limit of 5\e{8} $M_\odot$
        if its extent is as large as Figure~\ref{fig14} implies.

\subsection{Relation of the Hercules Thick Disk Cloud with the Hercules-Aquila Cloud}

The Hercules-Aquila Cloud is an overdensity or excess detected both above and
below the plane in SDSS DR5 data \citep{bel07}.
The authors identified a main sequence for stars with $20 < i < 22.5$ and $0.3 < g-i < 1.0$
by differencing Hess diagrams which contained the over-density with one that did not.
Using a color-magnitude relation for M92 they used an apparent turnoff color of 0.25 in $g-i$ and magnitudes between $19 < i < 20$ to infer a distance of 10-20 kpc for the cloud.
However, we think that there are several problems with this conclusion.

The star count excess described in this paper clearly exists in this direction for
stars with $16 < V < 18$ and $B - V < 0.6$.  Using the same color transformations used
for the photometric comparison with SDSS in \S {3}, we would expect to see our excess
appear  $17.5 < i < 19.5$ in their Figure 4, but it does not, despite its further identification in \citet{jur08} at the same magnitude range.
Furthermore, the width of the ridgeline in their Figure 4 is surprisingly small (0.4 dex in $g-i$).  Such a narrow range in color would imply that the Hercules-Aquila Cloud is exceedingly thin ($\sim 5000$ pc).  This is a hard dimension to reconcile with its projected width of 20 kpc and height of 15 kpc.  There is one more curious artifact in the image subtraction panel of their Figure 4.  Noise in a difference image should be expected to range between positive and negative values.  On their Figure 4 there is a curious all-negative locus directly redward of their indicated main sequence ridgeline.  This locus has a width of about 0.4 dex in $g-i$ and has very few positive S/N values within it.  This would be consistent with an oversubtraction in those magnitude and color ranges. The magnitude of its net 
significance 
approaches half that for their upper main sequence.  We posit that this oversubtraction hides the true distance to the overdensity.

\begin{figure}
\epsscale{0.8}
\plotone{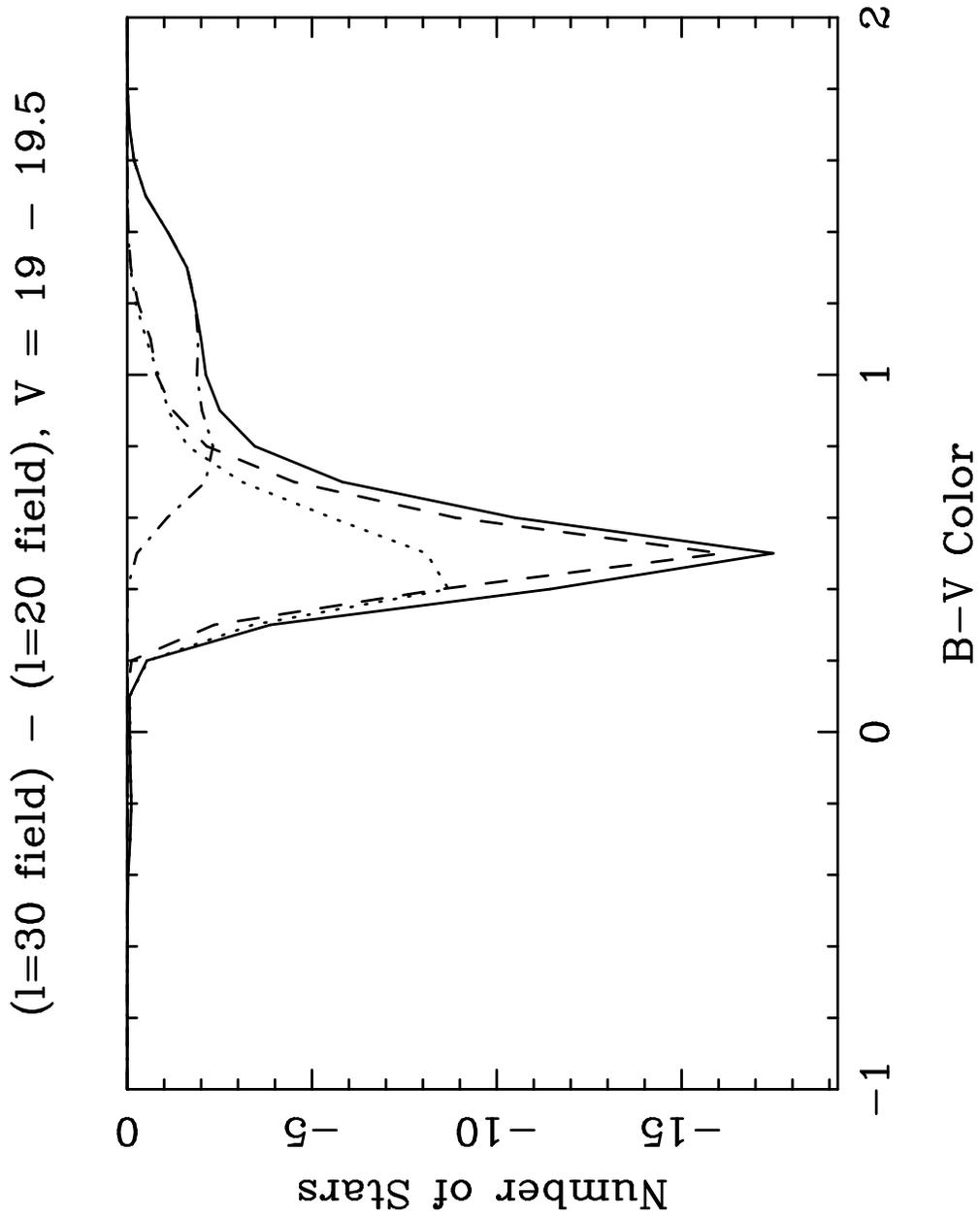}
\vspace{0.5in}
\caption{A GALMOD model of the subtraction performed in \cite{bel07} normalized to an area of 1 square degree.  The magnitudes and colors of the excess are strongly oversubtracted, leaving only the fainter and bluer portions of the excess as we detect it. \label{fig16}}
\end{figure}

SDSS does not cover the corresponding Q4 regions to allow for a direct comparison
at the  complementary {\it l} and {\it b}; therefore, \citet{bel07} removed the
large-scale effects of the stellar components of the galaxy by dividing the Hess diagrams
from an $8\degree \times 8 \degree$ ``on cloud" field at $l=30\degree, b=40\degree$ by a $16\degree \times 16\degree$ ``off cloud" field at $l=15\degree, b=40\degree$.  On the surface, this approach is very reasonable.  The scale height of the Disk limits its contribution at high latitudes and it's contribution does not greatly change with $l$.
Additionally, the Halo changes slowly as well considering a magnitude limited field.
However, the Thick Disk changes relatively rapidly between these two lines of sight and subtracting a lower longitude field causes an over-subtraction of these intermediate color stars. 
Figure ~\ref{fig16} shows a GALMOD recreation of the subtraction magnitudes similar to where \citet{bel07} place the turnoff of their M92 ridgeline fit.
The over-subtracted colors are in
the range $0.4 < B-V < 0.6$, corresponding to $0.4 < g - i < 0.8$, on the red side
of the M92 turnoff color of $g-i = 0.3$ determined in \citet{bel07}.  These Thick Disk stars
are more numerous in the ``off cloud" field than the overdensity would be in the ``on cloud" field, resulting in an oversubtraction which is apparent in their figure.
If their upper main sequence locus was actually 0.4 dex wider (the width of the oversubtraction artifact), the upper ridgeline on their
distance estimate would shift upwards approximately 2 magnitudes.  The Hercules-Aquila Cloud would shift much closer to us (minimum distance under 4 kpc) and would come into line with our estimates for the distance of the Hercules Cloud and the distance to the overdensity described in \citet{jur08}.

We propose that the Hercules-Aquila Cloud and the Hercules Thick Disk Cloud are the same feature, at the nearer distances of the Hercules Cloud.

\section{Discussion -- The Hercules Thick Disk Cloud in Context}

The star count asymmetry in Q1, the Hercules Cloud,  is associated with a Thick-Disk-like
density function having an upper boundary at $b\approx40\degree$.  The strength of the feature
decreases at lower latitudes, but then reappears below the plane, indicating that it is
not an isolated stream confined to one side of the Disk.  The photometric parallaxes for our  ``Blue'' population shows  the
star count excess extending along various sight lines from $\approx$ 1 to 6 kpc from the Sun.
The regions showing the strongest excess along these lines of sight have a very interesting
association with the increasing density in the direction of the bar in the Galactic 
plane and may have an associated  kinematic signature  \citep{paperIII}. Thus the  stars
participating in the over-density in Q1 may be either  members of a Thick Disk population
associated with the bar or result from a  dynamical response to the bar's passage.  From the calculations presented in Section 4.2 we infer a total mass for this feature which ranges from a conservative estimate of 2\e{6} $M_\odot$ if it is a 
local overdensity to  a potential upper limit of 5\e{8} $M_\odot$ if it is associated 
with the bar.  Finally, we argue that the Hercules-Aquila Cloud is much closer to the Sun than previously reported
and associated with the Hercules Thick Disk Cloud.

We have also identified an excess in a populations of  faint red stars in Q1, but much
closer to the Sun. It is possible that these stars may extend into the Solar neighborhood
and could be related to the local Hercules stream \citep{1998A&A...335L..61R, 1999ApJ...524L..35D, 2000AJ....119..800D, 2007ApJ...655L..89B, 2009IAUS..254..139W}.
This is certainly an interesting
possibility but one not easily resolved. The local moving groups are defined
by nearby bright stars distributed in a volume of space surrounding the Sun,
while in our approach to mapping the asymmetry with relatively faint stars,
direction is important.

Further work on this asymmetry in Q1, or the Hercules Cloud,  must concentrate
on several questions. First of all,  areas on the sky more than 1 square degree
must be observed along each line of sight to increase the statistical significance
of the detection.  Far more fields
in  Q4 and also below the plane and in other directions not covered by SDSS are needed for
Galactic structure studies in general, and also to improve our mapping of the excess
below the plane.  Large scale surveys at the lower latitudes may be required to statistically isolate the Hercules Cloud from the Galactic Disk.  
Does the Cloud or asymmetry extend towards  the Galactic center?
\citet{jur08} traced the over-density associated with the Cloud to $l=355\degree$, but it
disappears by $l=340\degree$ \citep{lar08}.  If it is associated with the bar
or due to a gravitational interaction with the bar, as we suspect, then the excess
would be expected to extend into Q4, but at much greater distances and fainter magnitudes. 
And finally, is the
asymmetry related to the local Hercules Stream which passes through the Solar neighborhood?

In Paper III, we discuss the kinematics of the associated stellar population 
and the possible origins of the Hercules Thick Disk Cloud.

\acknowledgments

This work was supported by  Collaborative National Science Foundation grants to Cabanela (AST0729989), Larsen (AST0507309) and Humphreys (AST0507170).  We  thank Steward Observatory and NOAO for observing support, and our respective home institutions for providing facilities support.  J.E. Cabanela thanks
undergraduate research assistants Joshua Swanson and Laura
Broaded for testing and reducing the original Y4KCam data
from 2006 April.  JAL would like to thank his students over the course of this project, Andrew Tucker and Aaron Haviland, for their work on modeling and data reduction and NRL grant N0001409WR40059 (FY09) for funds to support these student efforts.  He would also like to thank Debora Katz for observing assistance in May of 2006 at Steward Observatory.

{\it Facilities:} \facility{Bok (90Prime)}, \facility{CTIO:0.9m}, \facility{Blanco (Hydra)}, \facility{MMT (Hectospec)}.

\appendix

\section{GALMOD Predictions}

In this section we show the model-based  predictions for the  ratios in a symmetric galaxy 
with  parameters from \cite{lar03}.  Tables~\ref{tbl5}, \ref{tbl6} and \ref{tbl7} present the predictions for the ``Blue," ``Intermediate" and ``Red" color ranges defined in the text as a function of magnitude.  In addition, we compute the relative proportions of Disk/Halo/Thick Disk stars expected in each magnitude range.

\section{Count Ratio Results}

In this section we present the actual counts from the catalogs described in the text for all of 
the paired fields  across the lines of symmetry.  Tables~\ref{tbl8}, \ref{tbl9} and \ref{tbl10} present the actual count ratios together with the uncertainty and significance for stars from the ``Blue," ``Intermediate" and ``Red" color ranges defined in the text as a function of magnitude.

\textheight=7.5in

\clearpage
\begin{center}
\begin{deluxetable}{ccccccc}
\tabletypesize{\scriptsize}
\tablewidth{0pt}
\tablecaption{Field Observation Information for the Thick Disk Asymmetry Project.\label{tbl1}}
\tablehead{
\colhead{Field Name} &
\colhead{$l$} &
\colhead{$b$} &
\colhead{RA (J2000)} &
\colhead{Dec (J2000)} &
\colhead{Instrument} &
\colhead{Run Observed} 
}
\startdata
H020+20 & 19.96\arcdeg & 19.96\arcdeg & 17$^h$17$^m$12$^s$ & -01\arcdeg46'31 & 90Prime & 2006 May  \\ 
H020+32 & 20.01\arcdeg & 31.95\arcdeg & 16$^h$35$^m$28$^s$ & +04\arcdeg09'44 & 90Prime & 2006 May \\ 
H020+47 & 20.00\arcdeg & 46.99\arcdeg & 15$^h$42$^m$32$^s$ & +11\arcdeg17'18 & 90Prime & 2008 May \\ 
H020-47 & 20.00\arcdeg & -47.01\arcdeg & 21$^h$36$^m$27$^s$ & -28\arcdeg00'21 & Y4KCam & 2008 Oct \\ 
H023+40 & 23.00\arcdeg & 39.99\arcdeg & 16$^h$12$^m$01$^s$ & +09\arcdeg59'47 & 90Prime & 2008 May \\ 
H025+40 & 25.01\arcdeg & 39.96\arcdeg & 16$^h$14$^m$27$^s$ & +11\arcdeg28'33 & 90Prime & 2006 May \\ 
H027+40 & 27.00\arcdeg & 39.96\arcdeg & 16$^h$17$^m$24$^s$ & +12\arcdeg44'26 & 90Prime & 2008 May \\ 
H027+37 & 27.01\arcdeg & 36.99\arcdeg & 16$^h$28$^m$36$^s$ & +11\arcdeg28'52 & 90Prime & 2008 May \\
\enddata
\end{deluxetable}
\end{center}

\clearpage
\begin{deluxetable}{cccc}
\tablewidth{0pt}
\tabletypesize{\scriptsize}
\tablecaption{Photometric Zero Point Comparison with SDSS DR7.\label{tbl2}}
\tablehead{
\colhead{Field Name} &
\colhead{V Zero Point Difference} & 
\colhead{B-V Zero Point Difference} & 
\colhead{V-R Zero Point Difference} 
}
\startdata
H020+47 & 0.01 & 0.00 & -0.01 \\
H023+40 & -0.01 & -0.03 & -0.04 \\
H025+40 & -0.09 & 0.10 & -0.17 \\
H027+37 & -0.10 & 0.01 & -0.02 \\
H027+40 & -0.06 & -0.08 & -0.02 \\
H030+20 & -0.05 & 0.01 & 0.00 \\
H030-20 & -0.06 & -0.03 & 0.05 \\
H033+40 & 0.00 & -0.01 & -0.02 \\
H035+32 & -0.03 & 0.01 & 0.00 \\
H042+40 & -0.07 & 0.01 & -0.07 \\
H044+40 & 0.09 & -0.21 & -0.16 \\
H046+45 & -0.02 & -0.07 & 0.01 \\
H048+45 & -0.13 & 0.02 & -0.02 \\
H050+31 & -0.05 & -0.04 & -0.02 \\
H050-31 & -0.06 & 0.02 & -0.05 \\
H053+42 & 0.06 & -0.19 & -0.16 \\
H055+42 & -0.08 & 0.01 & -0.05 \\
\tableline
Average & $-0.04 \pm 0.05$ & $-0.03 \pm 0.08 $ & $-0.04 \pm 0.06$ \\
\enddata
\end{deluxetable}

\clearpage
\begin{deluxetable}{ccccccccccccccc}
\rotate
\setlength{\tabcolsep}{0.02in}
\tablewidth{0pt}
\tabletypesize{\scriptsize}
\tablecaption{Sample entries from the electronic catalog for of the Thick Disk Asymmetry Project.\label{tbl3}}
\tablehead{
\colhead{Star Name} &
\colhead{RA (J2000)} &
\colhead{Dec (J2000)} &
\colhead{$\sigma_{RA}$} &
\colhead{$\sigma_{Dec}$} &
\colhead{Class} &
\colhead{Tests} &
\colhead{$V$} &
\colhead{$\sigma_{V}$} &
\colhead{$U-B$} &
\colhead{$\sigma_{U-B}$} &
\colhead{$B-V$} &
\colhead{$\sigma_{B-V}$} &
\colhead{$V-R$} &
\colhead{$\sigma_{V-R}$}
}
\startdata
H020+20.000194 & 258.718077 & -01.600187&  0.000076&  0.000026&  00.00 & 4 & 21.17 & 00.05&  00.54&  00.16&  00.93&  00.08&  01.10&  00.06 \\
H020+20.000195 & 258.718108 & -02.002500&  0.000063&  0.000135&  00.40 & 3 & 21.91 & 00.16&  99.00&  99.00&  99.00&  99.00&  01.70&  00.17 \\
H020+20.000196 & 258.718169 & -02.080991&  0.000018&  0.000039&  00.20 & 4 & 20.61 & 00.07&  99.00&  99.00&  01.65&  00.14&  00.58&  00.08 \\
H020+20.000197 & 258.718219 & -01.711667&  0.000042&  0.000043 & 00.00 & 4 & 22.15 & 00.16&  99.00&  99.00&  99.00&  99.00&  01.33&  00.19 \\
H020+20.000198 & 258.718250 & -02.247588&  0.000032&  0.000073 & 00.05 & 4 & 22.64 & 00.22&  99.00&  99.00&  00.12&  00.27&  00.66&  00.27 \\
H020+20.000199 & 258.718303 & -02.239392&  0.000052&  0.000072 & 00.00 & 4 & 19.74 & 00.02&  00.12&  00.09&  01.33&  00.05&  01.00&  00.02 \\
\enddata
\end{deluxetable}

\begin{center}
\begin{deluxetable}{ccccccc}
\tabletypesize{\scriptsize}
\tablewidth{0pt}
\tablecaption{Area, Completeness and Blue Ridgeline Locations for the Thick Disk Asymmetry Project Fields.\label{tbl4}}
\tablehead{
\colhead{Field Name} &
\colhead{Field Area} &
\colhead{Completeness} &
\colhead{$(B-V)_{P1}$\tablenotemark{a}} &
\colhead{$(B-V)_{P2}$\tablenotemark{b}} &
\colhead{$(B-V)_{P3}$\tablenotemark{c}} &
\colhead{$(B-V)_{P4}$\tablenotemark{d}}
}
\startdata
H020+20 & 1.015 & 19.5 & 0.79 & 0.79 & 0.79 & 0.79 \\
H020+32 & 1.007 & 21.0 & 0.60 & 0.60 & 0.60 & 0.60 \\
H020+47 & 0.756 & 21.5 & 0.65 & 0.65 & 0.65 & 0.65 \\
H020-47 & 0.811 & 19.5 & 0.67 & 0.67 & 0.67 & 0.67 \\
H023+40 & 0.756 & 21.5 & 0.64 & 0.64 & 0.64 & 0.64 \\
H025+40 & 1.002 & 21.0 & 0.56 & 0.56 & 0.56 & 0.56 \\
H027+40 & 0.763 & 21.0 & 0.67 & 0.67 & 0.67 & 0.67 \\
H027+37 & 0.779 & 20.5 & 0.63 & 0.63 & 0.63 & 0.63 \\
\enddata
\tablenotetext{a}{~ Location of blue ridge for $16 < V < 18$.}
\tablenotetext{b}{~ Location of blue ridge for $18 < V < 19$.}
\tablenotetext{c}{~ Location of blue ridge for $19 < V < 20$.}
\tablenotetext{d}{~ Location of blue ridge for $20 < V < 21$.}
\end{deluxetable}
\end{center}

\begin{deluxetable}{cccccccccccc}
\tabletypesize{\scriptsize}
\rotate
\tablecaption{Galmod predictions for ``Blue" star counts with $0 < B-V < (B-V)_{P}$.\label{tbl5}}
\tablewidth{0pt}
\setlength{\tabcolsep}{0.02in}
\tablehead{
Fields & Joint & \multicolumn{5}{c}{GALMOD Ratio Predictions} & \multicolumn{5}{c}{Disk/Halo/Thick Disk Fraction of Total Stars} \\
Compared & Limit &  All & $16-18$ & $18-19$ & $19-20$ & $20-21$ & All & $16-18$ & $18-19$ & $19-20$ & $20-21$ \\
}
\startdata
\multicolumn{11}{l}{Quadrant 1/Quadrant 4 ratios above the Galactic Plane}\\ \tableline
H020+20/H340+20 & 18.0 & 1.00 & 1.00 & - & - & - & 0.44/0.23/0.33 & 0.44/0.23/0.33 & -/-/- & -/-/- & -/-/- \\
H020+32/H340+32 & 20.0 & 1.00 & 1.00 & 1.00 & 1.00 & - & 0.05/0.48/0.47 & 0.31/0.26/0.43 & 0.00/0.64/0.36 & 0.00/0.72/0.27 & -/-/- \\
H020+47/H340+47 & 21.0 & 1.00 & 1.00 & 1.00 & 1.00 & 1.00 & 0.04/0.55/0.41 & 0.06/0.41/0.53 & 0.00/0.71/0.28 & 0.01/0.81/0.18 & 0.01/0.59/0.40 \\
H023+40/H337+40 & 20.5 & 1.00 & 1.00 & 1.00 & 1.00 & 1.00 & 0.03/0.56/0.41 & 0.09/0.38/0.53 & 0.00/0.68/0.32 & 0.01/0.78/0.21 & 0.01/0.58/0.42 \\
H025+40/H335+40 & 21.0 & 1.00 & 1.00 & 1.00 & 1.00 & 1.00 & 0.06/0.51/0.43 & 0.09/0.38/0.53 & 0.00/0.68/0.31 & 0.01/0.78/0.21 & 0.01/0.58/0.42 \\
H027+37/H333+37 & 20.0 & 1.00 & 1.00 & 1.00 & 1.00 & - & 0.03/0.51/0.45 & 0.25/0.28/0.47 & 0.00/0.67/0.33 & 0.01/0.77/0.23 & -/-/- \\
H027+40/H333+40 & 20.5 & 1.00 & 1.00 & 1.00 & 1.00 & 1.00 & 0.03/0.57/0.41 & 0.21/0.30/0.49 & 0.00/0.68/0.31 & 0.01/0.79/0.21 & 0.01/0.58/0.41 \\
\enddata
\end{deluxetable}

\begin{deluxetable}{cccccccccccc}
\tabletypesize{\scriptsize}
\rotate
\tablecaption{Galmod predictions for ``Intermediate" stars with $(B-V)_{P} - 0.5 < B-V < (B-V)_{P} + 0.5$.\label{tbl6}}
\tablewidth{0pt}
\setlength{\tabcolsep}{0.02in}
\tablehead{
Fields & Joint &  \multicolumn{5}{c}{GALMOD Ratio Predictions} & \multicolumn{5}{c}{Disk/Halo/Thick Disk Fraction of Total Stars} \\
Compared &Limit&  All & $16-18$ & $18-19$ & $19-20$ & $20-21$ & All & $16-18$ & $18-19$ & $19-20$ & $20-21$ \\
}

\startdata
\multicolumn{11}{l}{Quadrant 1/Quadrant 4 ratios above the Galactic Plane}\\ \tableline
H020+20/H340+20 & 18.0 & 1.00 & 1.00 & - & - & - & 0.67/0.10/0.22 & 0.67/0.10/0.22 & -/-/- & -/-/- & -/-/- \\
H020+32/H340+32 & 20.0 & 1.00 & 1.00 & 1.00 & 1.00 & - & 0.29/0.28/0.42 & 0.55/0.14/0.30 & 0.23/0.30/0.47 & 0.07/0.41/0.51 & -/-/- \\
H020+47/H340+47 & 21.0 & 1.00 & 1.00 & 1.00 & 1.00 & 1.00 & 0.17/0.40/0.43 & 0.42/0.20/0.38 & 0.10/0.39/0.51 & 0.02/0.51/0.47 & 0.38/0.25/0.37 \\
H023+40/H337+40 & 20.5 & 1.00 & 1.00 & 1.00 & 1.00 & 1.00 & 0.20/0.36/0.44 & 0.47/0.18/0.35 & 0.14/0.35/0.50 & 0.03/0.47/0.50 & 0.30/0.27/0.43 \\
H025+40/H335+40 & 21.0 & 1.00 & 1.00 & 1.00 & 1.00 & 1.00 & 0.20/0.37/0.43 & 0.47/0.18/0.35 & 0.14/0.35/0.50 & 0.03/0.47/0.49 & 0.31/0.27/0.43 \\
H027+37/H333+37 & 20.0 & 1.00 & 1.00 & 1.00 & 1.00 & - & 0.26/0.31/0.43 & 0.52/0.16/0.32 & 0.17/0.33/0.49 & 0.04/0.46/0.50 & -/-/- \\
H027+40/H333+40 & 20.5 & 1.00 & 1.00 & 1.00 & 1.00 & 1.00 & 0.20/0.36/0.44 & 0.50/0.17/0.33 & 0.15/0.35/0.50 & 0.03/0.48/0.49 & 0.31/0.27/0.43 \\
\enddata
\end{deluxetable}

\begin{deluxetable}{cccccccccccc}
\tabletypesize{\scriptsize}
\rotate
\tablecaption{Galmod predictions for ``Red" stars with $(B-V)_{P} + 0.5 < B-V < (B-V)_{P} + 1.5$.\label{tbl7}}
\tablewidth{0pt}
\setlength{\tabcolsep}{0.02in}
\tablehead{
Fields & Joint &  \multicolumn{5}{c}{GALMOD Ratio Predictions} & \multicolumn{5}{c}{Disk/Halo/Thick Disk Fraction of Total Stars} \\
Compared &Limit &  All & $16-18$ & $18-19$ & $19-20$ & $20-21$ & All & $16-18$ & $18-19$ & $19-20$ & $20-21$ \\
}

\startdata
\multicolumn{11}{l}{Quadrant 1/Quadrant 4 ratios above the Galactic Plane}\\ \tableline
H020+20/H340+20 & 18.0 & 1.00 & 1.00 & - & - & - & 0.99/0.00/0.01 & 0.99/0.00/0.01 & -/-/- & -/-/- & -/-/- \\
H020+32/H340+32 & 20.0 & 1.00 & 1.00 & 1.00 & 1.00 & - & 0.94/0.01/0.05 & 0.98/0.00/0.01 & 0.95/0.01/0.04 & 0.92/0.01/0.07 & -/-/- \\
H020+47/H340+47 & 21.0 & 1.00 & 1.00 & 1.00 & 1.00 & 1.00 & 0.90/0.02/0.08 & 0.98/0.00/0.02 & 0.93/0.01/0.06 & 0.87/0.02/0.11 & 1.00/0.00/0.00 \\
H023+40/H337+40 & 20.5 & 1.00 & 1.00 & 1.00 & 1.00 & 1.00 & 0.89/0.02/0.09 & 0.98/0.00/0.02 & 0.94/0.01/0.05 & 0.89/0.02/0.09 & 1.00/0.00/0.00 \\
H025+40/H335+40 & 21.0 & 1.00 & 1.00 & 1.00 & 1.00 & 1.00 & 0.92/0.01/0.07 & 0.98/0.00/0.02 & 0.94/0.01/0.05 & 0.89/0.02/0.09 & 1.00/0.00/0.00 \\
H027+37/H333+37 & 20.0 & 1.00 & 1.00 & 1.00 & 1.00 & - & 0.93/0.01/0.06 & 0.98/0.00/0.02 & 0.95/0.01/0.05 & 0.90/0.02/0.08 & -/-/- \\
H027+40/H333+40 & 20.5 & 1.00 & 1.00 & 1.00 & 1.00 & 1.00 & 0.89/0.02/0.09 & 0.98/0.00/0.02 & 0.94/0.01/0.05 & 0.89/0.02/0.09 & 1.00/0.00/0.00 \\
\enddata
\end{deluxetable}

\begin{deluxetable}{cccccccccccc}
\tabletypesize{\scriptsize}
\rotate
\tablecaption{Star counts super-ratio for ``Blue" stars with $0 < B-V < (B-V)_{P}$.\label{tbl8}}
\tablewidth{0pt}
\tablehead{
Fields & Joint & \multicolumn{2}{c}{All Data} & \multicolumn{2}{c}{$ 16 < V < 18$} & \multicolumn{2}{c}{$ 18 < V < 19$} & \multicolumn{2}{c}{$ 19 < V < 20$} &  \multicolumn{2}{c}{$ 20 < V < 21$} \\
Compared &  Limit & Ratio & $s$ & Ratio & $s$ & Ratio & $s$ & Ratio & $s$ & Ratio & $s$ \\
}
\startdata
\multicolumn{11}{l}{Quadrant 1/Quadrant 4 ratios above the Galactic Plane}\\ \tableline
H020+20/H340+20 & 18.0 & $ 0.97 \pm 0.06 $ & 0.50 &  $ 0.97 \pm 0.06 $ & 0.50 &  $ - \pm - $ & - &  $ - \pm - $ & - &  $ - \pm - $ & - \\
H020+32/H340+32 & 20.0 & $ 1.04 \pm 0.07 $ & 0.57 &  $ 1.06 \pm 0.07 $ & 0.86 &  $ 0.96 \pm 0.08 $ & 0.48 &  $ 1.09 \pm 0.07 $ & 1.27 &  $ - \pm - $ & - \\
H020+47/H340+47 & 21.0 & $ 0.98 \pm 0.06 $ & 0.30 &  $ 1.13 \pm 0.14 $ & 0.91 &  $ 1.09 \pm 0.12 $ & 0.77 &  $ 0.84 \pm 0.08 $ & 1.86 &  $ 0.94 \pm 0.08 $ & 0.78 \\
H023+40/H337+40 & 20.5 & $ 1.05 \pm 0.07 $ & 0.79 &  $ 1.20 \pm 0.12 $ & 1.61 &  $ 1.04 \pm 0.11 $ & 0.39 &  $ 0.96 \pm 0.09 $ & 0.39 &  $ 0.98 \pm 0.10 $ & 0.23 \\
H025+40/H335+40 & 21.0 & $ 1.01 \pm 0.09 $ & 0.14 &  $ 1.10 \pm 0.07 $ & 1.40 &  $ 1.01 \pm 0.13 $ & 0.07 &  $ 0.98 \pm 0.10 $ & 0.24 &  $ 0.98 \pm 0.08 $ & 0.26 \\
H027+37/H333+37 & 20.0 & $ 1.02 \pm 0.07 $ & 0.24 &  $ 1.16 \pm 0.12 $ & 1.35 &  $ 1.05 \pm 0.10 $ & 0.56 &  $ 0.86 \pm 0.06 $ & 2.24 &  $ - \pm - $ & - \\
H027+40/H333+40 & 20.5 & $ 1.09 \pm 0.08 $ & 1.17 &  $ 1.41 \pm 0.28 $ & 1.46 &  $ 1.07 \pm 0.11 $ & 0.63 &  $ 0.97 \pm 0.09 $ & 0.28 &  $ 0.87 \pm 0.10 $ & 1.30 \\
\enddata
\end{deluxetable}

\begin{deluxetable}{cccccccccccc}
\tabletypesize{\scriptsize}
\rotate
\tablecaption{Star count super-ratio for ``Intermediate" stars with $(B-V)_{P} - 0.5 < B-V < (B-V)_{P} + 0.5$.\label{tbl9}}
\tablewidth{0pt}
\tablehead{
Fields & Joint & \multicolumn{2}{c}{All Data} & \multicolumn{2}{c}{$ 16 < V < 18$} & \multicolumn{2}{c}{$ 18 < V < 19$} & \multicolumn{2}{c}{$ 19 < V < 20$} &  \multicolumn{2}{c}{$ 20 < V < 21$} \\
Compared &  Limit & Ratio & $s$ & Ratio & $s$ & Ratio & $s$ & Ratio & $s$ & Ratio & $s$ \\
}
\startdata
\multicolumn{11}{l}{Quadrant 1/Quadrant 4 ratios above the Galactic Plane}\\ \tableline
H020+20/H340+20 & 18.0 & $ 0.98 \pm 0.02 $ & 0.91 &  $ 0.98 \pm 0.02 $ & 0.91 &  $ - \pm - $ & - &  $ - \pm - $ & - &  $ - \pm - $ & - \\
H020+32/H340+32 & 20.0 & $ 0.96 \pm 0.02 $ & 2.34 &  $ 0.93 \pm 0.05 $ & 1.63 &  $ 0.96 \pm 0.03 $ & 1.16 &  $ 0.98 \pm 0.03 $ & 0.48 &  $ - \pm - $ & - \\
H020+47/H340+47 & 21.0 & $ 1.08 \pm 0.03 $ & 2.89 &  $ 1.05 \pm 0.07 $ & 0.67 &  $ 1.11 \pm 0.07 $ & 1.61 &  $ 1.00 \pm 0.05 $ & 0.03 &  $ 1.14 \pm 0.05 $ & 3.01 \\
H023+40/H337+40 & 20.5 & $ 1.13 \pm 0.03 $ & 4.29 &  $ 1.20 \pm 0.08 $ & 2.35 &  $ 1.18 \pm 0.06 $ & 2.82 &  $ 1.06 \pm 0.05 $ & 1.15 &  $ 1.07 \pm 0.06 $ & 1.14 \\
H025+40/H335+40 & 21.0 & $ 1.02 \pm 0.02 $ & 0.72 &  $ 1.03 \pm 0.04 $ & 0.68 &  $ 0.96 \pm 0.05 $ & 0.83 &  $ 1.04 \pm 0.04 $ & 0.96 &  $ 1.02 \pm 0.04 $ & 0.49 \\
H027+37/H333+37 & 20.0 & $ 0.96 \pm 0.03 $ & 1.41 &  $ 1.00 \pm 0.05 $ & 0.08 &  $ 0.96 \pm 0.05 $ & 0.76 &  $ 0.93 \pm 0.04 $ & 1.88 &  $ - \pm - $ & - \\
H027+40/H333+40 & 20.5 & $ 1.04 \pm 0.03 $ & 1.28 &  $ 1.12 \pm 0.09 $ & 1.28 &  $ 1.02 \pm 0.06 $ & 0.41 &  $ 0.99 \pm 0.05 $ & 0.12 &  $ 1.00 \pm 0.06 $ & 0.04 \\
\enddata
\end{deluxetable}

\begin{deluxetable}{cccccccccccc}
\tabletypesize{\scriptsize}
\rotate
\tablecaption{Star count super-ratio for ``Red" stars with $(B-V)_{P} + 0.5 < B-V < (B-V)_{P} + 1.5$.\label{tbl10}}
\tablewidth{0pt}
\tablehead{
Fields & Joint & \multicolumn{2}{c}{All Data} & \multicolumn{2}{c}{$ 16 < V < 18$} & \multicolumn{2}{c}{$ 18 < V < 19$} & \multicolumn{2}{c}{$ 19 < V < 20$} &  \multicolumn{2}{c}{$ 20 < V < 21$} \\
Compared &  Limit & Ratio & $s$ & Ratio & $s$ & Ratio & $s$ & Ratio & $s$ & Ratio & $s$ \\
}
\startdata
\multicolumn{11}{l}{Quadrant 1/Quadrant 4 ratios above the Galactic Plane}\\ \tableline
H020+20/H340+20 & 18.0 & $ 0.74 \pm 0.06 $ & 4.46 &  $ 0.74 \pm 0.06 $ & 4.46 &  $ - \pm - $ & - &  $ - \pm - $ & - &  $ - \pm - $ & - \\
H020+32/H340+32 & 20.0 & $ 0.97 \pm 0.04 $ & 0.73 &  $ 0.93 \pm 0.10 $ & 0.69 &  $ 0.96 \pm 0.07 $ & 0.66 &  $ 0.99 \pm 0.05 $ & 0.17 &  $ - \pm - $ & - \\
H020+47/H340+47 & 21.0 & $ 1.05 \pm 0.04 $ & 1.25 &  $ 1.30 \pm 0.27 $ & 1.09 &  $ 1.20 \pm 0.11 $ & 1.81 &  $ 1.06 \pm 0.07 $ & 0.84 &  $ 0.98 \pm 0.05 $ & 0.38 \\
H023+40/H337+40 & 20.5 & $ 1.23 \pm 0.06 $ & 4.09 &  $ 1.12 \pm 0.19 $ & 0.61 &  $ 1.15 \pm 0.10 $ & 1.40 &  $ 1.32 \pm 0.08 $ & 3.84 &  $ 1.21 \pm 0.08 $ & 2.45 \\
H025+40/H335+40 & 21.0 & $ 1.00 \pm 0.03 $ & 0.03 &  $ 0.92 \pm 0.14 $ & 0.55 &  $ 0.91 \pm 0.07 $ & 1.22 &  $ 1.11 \pm 0.06 $ & 1.67 &  $ 0.98 \pm 0.04 $ & 0.45 \\
H027+37/H333+37 & 20.0 & $ 0.92 \pm 0.05 $ & 1.45 &  $ 0.92 \pm 0.11 $ & 0.73 &  $ 1.07 \pm 0.10 $ & 0.72 &  $ 0.86 \pm 0.06 $ & 2.58 &  $ - \pm - $ & - \\
H027+40/H333+40 & 20.5 & $ 1.13 \pm 0.04 $ & 3.14 &  $ 1.06 \pm 0.14 $ & 0.40 &  $ 1.18 \pm 0.10 $ & 1.86 &  $ 1.13 \pm 0.07 $ & 1.97 &  $ 1.13 \pm 0.07 $ & 1.89 \\
\enddata
\end{deluxetable}

\clearpage

\clearpage

\clearpage

\clearpage

\clearpage

\clearpage

\clearpage

\clearpage

\clearpage

\clearpage

\clearpage

\clearpage

\clearpage

\end{document}